\documentclass[prc,nobibnotes,preprint,superscriptaddress,showpacs,nofootinbib,tightenlines,amsmath]{revtex4-1}
\usepackage{graphics}
\usepackage{epsfig}
\usepackage{slashed}
\usepackage{pbox,calc}
\usepackage{braket}
\usepackage{bbm}
\usepackage{amsmath, amsfonts, amssymb}

\usepackage{times}
\usepackage{tabularx}

\usepackage{xcolor}
\definecolor{frenchblue}{rgb}{0.0, 0.45, 0.73}
\definecolor{RoyalRed}{RGB}{157, 16, 45}

\usepackage{bigints}
\begin{document}
\title{\textbf{Electric dipole moments of charm baryons using dimension-six operators}}
\author{Y. \"Unal}
\email{yuenalsa@gmail.com \\
yaseminunal@comu.edu.tr}
\affiliation{Physics Department, \c{C}anakkale  Onsekiz Mart University, 17100 \c{C}anakkale, Turkey}
%\date{April 7, 2021}
%-------------------------------------------------				

\begin{abstract}
We investigate the $C\!P$-odd electric dipole moments (EDMs) of spin-1/2 charm baryons considering $C\!P$-violating dimension-6 operators in the Standard Model effective field theory.  In the framework of heavy-baryon chiral perturbation theory, we  calculate the EDMs of single-charm baryons and present the estimates for beyond-the-standard model physics appearing at the TeV scale. 
\end{abstract}

%\keywords{Heavy baryon chiral perturbation theory \and $C\!P$ violation \and Electric dipole moment}
\maketitle

\section{Introduction}

The $CP$ transformation combines the parity transformation ($P$), the inversion of space, and charge conjugation ($C$) which interchanges particles with their antiparticles. The phenomenon of $C\!P$ violation is the breaking of the
combination of charge-conjugation symmetry and parity symmetry. 
Nature was considered symmetrical under these transformations until the first evidence of $C\!P$ violation in 1964~\cite{Christenson:1964fg}. Even though the SM of particle physics includes CP Violation, it is widely believed that it does not offer a valid mechanism for generating the matter-antimatter asymmetry, the so-called baryon asymmetry of the universe~\cite{Riotto:1999yt, Sakharov:1967dj}. One of the reasons for that is the amount of $C\!P$ violation produced in the SM is not large enough to explain the matter-antimatter imbalance~\cite{Canetti:2012zc}. Searching for signals of $C\!P$ violation in many other processes might help thus identify potential sources of this violation beyond the SM.

Electric dipole moment experiments are among the most sensitive probes of $C\!P$ violating physics beyond the standard model.  EDMs of stable systems containing light quarks have been highly considered~\cite{deVries:2010ah, deVries:2012ab, Yamanaka:2020kjo, Kley:2021yhn}. In recent years, however, various experimental studies focussed on searching for EDMs of baryons including heavier quarks, mostly triggered by ongoing plans for dedicated experiments to measure EDMs of heavy baryons~\cite{Fomin:2017ltw,Aiola:2020yam,Bagli:2017foe, Merli:2022lqf, Biryukov:2021cml, Botella:2016ksl}. 
Since there has not been much theoretical work to accompany these progressive experimental studies, we address the analysis of heavy charmed baryon EDMs in this article. Moreover, this work is an extension of the calculation in~\cite{Unal:2020ezc} to $C\!P$-odd sources beyond the QCD $\theta$-term including charm quarks with dimension-6 $C\!P$-violating operators in the SMEFT. 

The experimental EDM programs are mostly concentrated on the measurements of the hadrons involving light quarks so far because the measurements of the heavy quark EDMs are difficult due to their short lifetimes (see, e.g., Refs. \cite{Dekens:2014jka, Braaten:1990zt,Chien:2015xha,Gisbert:2019ftm,Haisch:2021hcg}. Indirect limits on charm and beauty quark EDM, which indirectly constrain $C\!P$-violating effects, are set from different experimental measurements~\cite{Ema:2022pmo, Sala:2013osa, Zhao:2016jcx, Cordero-Cid:2007cmf}. For instance, an indirect bound on the charm quark EDM is derived from the experimental limit on the neutron EDM~\cite{Ema:2022pmo}. However, it has been observed that the experimental limits on EDMs involving heavier quarks are much stronger than the same experiments containing light quarks. Our goal thus is to obtain EDMs of charm baryons that could give more direct information on $C\!P$-violating interactions involving heavy quarks. For that purpose, we set up an EFT to calculate these contributions in a systematic fashion combining Chiral Perturbation Theory (ChPT) and heavy-quark EFT.

The techniques developed in~\cite{Unal:2021lhb} for the calculation of bottom baryon EDMs have been applied in the present article to calculate EDMs of charmed baryons. Throughout this work, we shall very often use the convenient language of it.
The article is organized as follows. In Sec. II, we introduce dimension-6 $C\!P$-violating SMEFT operators involving charm quarks. In Sec. III, we present these operators at the hadronic level using chiral perturbation theory. In Sec. IV, we calculate the EDMs of charm quark baryons at leading order for each source of $C\!P$ violation.  Sec. V contains the discussion of the expected magnitudes of EDMs.  In Sec. VI, we give the conclusion. Some technicalities are given in the Appendix.

\section{CP-violating operators including charm quarks}\label{dim6b}
In the following, we provide $C\!P$-violating operators at the quark level including light and charm quarks.  We stick to the notation of Ref. \cite{Unal:2021lhb}, where more details can be found.  In terms of the charm quark bilinears, the resulting $P$- and $T$-violating effective dimension-6 operators are given by~\cite{Grzadkowski:2010es, deVries:2012ab, Bsaisou:2014goa} 
\begin{equation}\
\begin{aligned}
\mathcal{L}_{c,  \mathrm{qEDM}}^{(6)}=~& d_c \bar{c}~ \sigma^{\mu \nu} \gamma_5 c ~F_{\mu \nu}\,, \\
\mathcal{L}_{c,  \mathrm{qCEDM}}^{(6)}=~& \tilde{d}_c \bar{c}~\sigma^{\mu \nu} \gamma_5 \lambda^a c~G_{\mu \nu}^a\,, 
\\
\mathcal{L}_{c,  \mathrm{4q}}^{(6)}=~& i \kappa_1^{uc} (\bar{u}u\bar{c}\gamma_5 c + \bar{u}\gamma_5u\bar{c}c -
\bar{c} \gamma_5 u \bar{u}c - \bar{c}u \bar{u} \gamma_5 c) + i \kappa_1^{dc} (\bar{d}d\bar{c}\gamma_5 c
+ \bar{d}\gamma_5d\bar{c}c 
\\
&- \bar{c} \gamma_5 d \bar{d}c - \bar{c}d \bar{d} \gamma_5 c) + i \kappa_1^{sc} (\bar{s}s\bar{c}
\gamma_5 c + \bar{s}\gamma_5s\bar{c}c - \bar{c} \gamma_5 s \bar{s}c - \bar{c}s \bar{s} \gamma_5 c)
 \\
                                                    &+i \kappa_8^{uc} (\bar{u} \lambda^a u\bar{c}\gamma_5 \lambda^a c + \bar{u}\gamma_5 \lambda^a u\bar{c} \lambda^a c - \bar{c} \gamma_5 \lambda^a u \bar{u} \lambda^a c - \bar{c} \lambda^a u \bar{u} \gamma_5 \lambda^a c) 
\\
                                                    &+i \kappa_8^{dc} (\bar{d} \lambda^a d\bar{c}\gamma_5 \lambda^a c + \bar{d}\gamma_5 \lambda^a d\bar{c} \lambda^a c - \bar{c} \gamma_5 \lambda^a d \bar{d} \lambda^a c - \bar{c} \lambda^a d \bar{d} \gamma_5 \lambda^a c) \\
                                                    &+i \kappa_8^{sc} (\bar{s} \lambda^a s\bar{c}\gamma_5 \lambda^a c + \bar{s}\gamma_5 \lambda^a s\bar{c} \lambda^a c - \bar{c} \gamma_5 \lambda^a s \bar{s} \lambda^a c - \bar{c} \lambda^a s \bar{s} \gamma_5 \lambda^a c)\,,
\\
	\mathcal{L}_{c,  \mathrm{4qLR}}^{(6)}=~& i \rho_{1}^{dc} V_{dc} (\bar{c}_R\gamma_{\mu} d_R \bar{d}_L\gamma^{\mu} c_L)-i\rho_{1}^{dc} V_{dc}^* (\bar{d}_R\gamma^{\mu} c_R \bar{c}_L\gamma_{\mu} d_L ) \\
                                                   &+i \rho_{8}^{dc} V_{dc} (\bar{c}_R \gamma_{\mu} \lambda^a d_R \bar{d}_L \gamma^{\mu} \lambda^a c_L)-i\rho_{8}^{dc} V_{dc}^* (\bar{d}_R\gamma^{\mu} \lambda^ a c_R \bar{c}_L\gamma_{\mu} \lambda^ a d_L ) \\
                                                   &+ i \rho_{1}^{sc} V_{sc} (\bar{c}_R\gamma_{\mu} s_R \bar{s}_L\gamma^{\mu} c_L)-i\rho_{1}^{sc} V_{sc}^* (\bar{s}_R\gamma^{\mu} c_R \bar{c}_L\gamma_{\mu} s_L ) \\
                                                   &+i \rho_{8}^{sc} V_{sc} (\bar{c}_R \gamma_{\mu} \lambda^a s_R \bar{s}_L \gamma^{\mu} \lambda^a c_L)-i\rho_{8}^{sc} V_{sc}^* (\bar{s}_R\gamma^{\mu} \lambda^ a c_R \bar{c}_L\gamma_{\mu} \lambda^ a s_L )\,,
	\label{pt_vio_op}
	\end{aligned}
\end{equation}
where $V_{dc}$ and $V_{sc}$ are elements of the CKM matrix, $F_{\mu \nu}$,
and $G_{\mu \nu}^a$ are the electromagnetic and the gluon field-strength tensors, respectively.  It should be noted here that only one CKM matrix element contributes to the $C\!P$-violating operators in the $b$-quark case, while two CKM matrix elements are allowed in the $c$-quark sector. 

In the SM-EFT Lagrangian, the qEDM and qCEDM operators are based on the following dimension-six operators 
\begin{eqnarray}
\begin{aligned}
\mathcal L_{\mathrm{4q}}= \,& C^{bB}\,(\bar Q_2 \sigma^{\mu\nu} c_{R_b}) H B_{\mu\nu} + C^{bW}\,(\bar Q_2 \sigma^{\mu\nu} \tau^a c_{R_b}) H W^a_{\mu\nu} \\
& + C^{bG}\,(\bar Q_2 \sigma^{\mu\nu} \lambda^a c_{R_b}) H G^a_{\mu\nu} + {\rm h.c.}
\end{aligned}
\label{eq:EFT4fermion}
\end{eqnarray}
where $Q_2$ denotes a left-doublet of second-generation quarks, $H$ is the Higgs doublet to preserve gauge invariance and
$B_{\mu\nu}$ and $W_{\mu\nu}^a$ are the field strength tensors of the $U(1)_Y$ and $SU(2)_L$ gauge groups, in order.
Below the electroweak symmetry-breaking scale of the
breakdown $SU(2)_L \times U(1)_Y\rightarrow U(1)_{em}$, the charm qEDM and the charm qCEDM result from dimension-six dipole operators. There are many models for physics BSM in which these dipole operators are generated (See, e.g., Refs.  \cite{Nakai:2016atk, Jung:2013hka, Dekens:2018bci}).
It is common to scale the dipoles with the heavy quark Yukawa in most of these models and thus we expect $d_c, \tilde{d}_c \sim m_c/\Lambda^2$.

The four-quark operators in $\mathcal{L}_{c,  \mathrm{4q}}^{(6)}$ are generated from the gauge-invariant operator of the form   
\begin{eqnarray}
\mathcal L_{\mathrm{4q}}=C^{abcd}_{\mathrm{4q}}\,(\bar Q_a^I  u_{R_b}) \epsilon_{IJ}(\bar Q_c^J  d_{R_d}) +{\rm h.c.}+\dots\,,
\label{eq:EFT4fermion}
\end{eqnarray}
where the ellipses indicate terms with the additional color structure, and $abcd$ are quark generation indices.
These operators
induce $\mathcal{L}_{c,  \mathrm{4q}}^{(6)}$ for the generation indices $a=b=\{1,2\}$ and $c=d=3$ or $a=d=3$
and $b=c=\{1,2\}$. 
The coupling constants are expected to scale as
$\kappa^{uc,dc,sc}_{1,8}\sim 1/\Lambda^2$. An example where the CP-odd four-quark operators are induced
can be seen in leptoquark models \cite{Dekens:2018bci}. 

The four-quark left-right operators in $\mathcal{L}_{c, 4qLR}^{(6)}$ are generated from the gauge-invariant operator of the form 
\begin{eqnarray}
\mathcal L_{\mathrm{4qLR}}=
C^{ab}_{\mathrm{4qLR}}\,\left(\tilde{H}^{\dagger} D_{\mu} H\right) \, \bar{d}^a_R \gamma^\mu  c^b_R
+ C^{ab}_{\mathrm{4qLR}}\,\left(\tilde{H}^{\dagger} D_{\mu} H\right) \, \bar{s}^a_R \gamma^\mu  c^b_R+{\rm h.c.}~.
\end{eqnarray}
The interactions in
$\mathcal{L}_{c,  4qLR}^{(6)}$ are generated at tree level between
quarks after electroweak symmetry breaking. Integrating out the $W$ bosons and Higgs fields the CKM elements of $V_{dc}$ and $V_{sc}$ are obtained. The contribution of higher-dimensional
operators from $\mathcal{L}_{c, 4qLR}^{(6)}$ will be proportional to $\rho_{1,8}^{qc} \sim
v^2/(m_W^2 \Lambda^2) \sim 1/\Lambda^2$, where $q = (d, s)$. The CP-odd four-quark operators can appear in the minimal
left-right symmetric model (See, e.g., \cite{Dekens:2021bro} for a recent EDM analysis).

\section{Chiral perturbation theory for charm baryons}\label{secChPT}

In this section, we list the effective interaction Lagrangians relevant for the calculation of the EDMs of the charmed baryons. To include heavy charm quarks into standard ChPT, we follow the same way which has been presented in \cite{Cheng:1993kp, Yan:1992gz}. The relevant $P$- and $T$-conserving free and interaction Lagrangians up to the second
chiral order in a covariant formalism are given by~\cite{Yan:1992gz,  Shi:2018rhk, Jiang:2014ena, Borasoy:2000pq} 
\begin{equation}
\begin{aligned}
  \mathcal{L}_{\phi}^{(2)}= ~& \frac{F_{\pi}^2}{4}\left\{\text{Tr}[D_\mu U(D^\mu U)^\dagger]+\text{Tr}(\chi U^\dagger+U \chi^\dagger)\right\},\\
  \mathcal{L}_{\text{free}}^{(1)}= ~& \frac{1}{2} \langle \bar{B}_{\bar{3}}(i \slashed{D}-m_{\bar{3}})
  B_{\bar{3}} \rangle
+\langle \bar{B}_{6}(i \slashed{D}-m_{6}) B_{6} \rangle\,,\\                                                                                                                
\mathcal{L}_{\text{int}}=~ ~ &\frac{h_1}{2} \langle \bar{B}_{6}\slashed{u} \gamma_5 {B}_{6} \rangle+\frac{h_2}{2} \langle \bar{B}_{6}\slashed{u} \gamma_5 {B}_{\bar{3}}+h.c. \rangle+\frac{h_3}{2} \langle \bar{B}_{\bar{3}}\slashed{u} \gamma_5 {B}_{\bar{3}} \rangle \,,  \\                                                                                                                                                    
\mathcal{L}_{B \gamma}^{(2)}= ~& \alpha_1 \langle \bar{B}_{\bar{3}} \sigma^{\mu \nu}
F_{\mu \nu}^+B_{\bar{3}} \rangle
+\alpha_2\langle \bar{B}_{6} \sigma^{\mu \nu} F_{\mu \nu}^+ B_{6} \rangle + \alpha_3 \langle \bar{B}_{6} \sigma^{\mu \nu} F_{\mu \nu}^+B_{3}
+ h.c. \rangle+\alpha_4 \langle \bar{B}_{\bar{3}} \sigma^{\mu \nu} B_{\bar{3}} \rangle \langle F_{\mu \nu}^+\rangle \\
&+\alpha_5 \langle \bar{B}_{6} \sigma^{\mu \nu} B_{6} \rangle\langle F_{\mu \nu}^+\rangle \,,
\label{Meson-baryon Lagr}				    
\end{aligned}
\end{equation} 
where $F_\pi$ is the pion-decay constant in the chiral limit, $B_{\bar{3}}$ and $B_{6}$ represent the spin-1/2 anti-symmetric triplet and symmetric sextet charm baryon states in the SU(3) flavor representation given by the following matrices,  respectively, ~    
\begin{equation}
  	B_{\bar{3}}=
  \begin{pmatrix}
               0                    & \Lambda_{c}^+        & \Xi_{c}^+ \\
    - \Lambda_{c}^+      &        0                       & \Xi_{c}^0 \\
    - \Xi_{c}^+                & -\Xi_{c}^0                &     0
  \end{pmatrix},\quad  	
  	B_6=
  \begin{pmatrix}
    \Sigma_{c}^{++}                           & \frac{\Sigma_{c}^+}{\sqrt{2}}           & \frac{\Xi_{c}^{'+}}{\sqrt{2}}  \\
    \frac{\Sigma_{c}^{+}}{\sqrt{2}}    & \Sigma_{c}^0                                  & \frac{\Xi_{c}^{'0}}{\sqrt{2}}  \\
    \frac{\Xi_{c}^{'+}}{\sqrt{2}}           & \frac{\Xi_{c}^{'0}} {\sqrt{2}}           & \Omega_{c}^{0}
  \end{pmatrix} .
	\end{equation}
	%--
	The Goldstone boson octet is denoted by 
\begin{equation}
	\phi=
	\begin{pmatrix}
	\frac{1}{\sqrt{2}}\pi^0+ \frac{1}{\sqrt{6}}\eta                      & \pi^+        & K^+ \\
	\pi^-      &        -\frac{1}{\sqrt{2}}\pi^0+ \frac{1}{\sqrt{6}}\eta                        & K^0 \\
	K^-                & \bar{K}^0                &     - \frac{2}{\sqrt{6}}\eta       
	\end{pmatrix}\,.
	\end{equation} 
	%
%--
The Lagrangians in Eq.~(\ref{Meson-baryon Lagr}) are constructed in terms of the following building blocks 
\begin{equation}
\begin{aligned}
	D_{\mu}^{} B = &\; \partial_{\mu}^{} B + \Gamma_{\mu}^{} B + B \Gamma_{\mu}^{T} ,\\
	\Gamma_{\mu} =& \;\frac{1}{2} \left[ u^{\dagger} ( \partial_{\mu} - i r_{\mu} ) u + u ( \partial_{\mu} - i l_{\mu} ) u^{\dagger} \right] ,\\
	F_{\mu \nu}^+ =&\; u^{\dagger} Q_h F_{\mu \nu} u + u Q_h F_{\mu \nu} u^{\dagger} ,\\
	u_{\mu} =&\; i \left[ u^{\dagger} ( \partial_{\mu} - i r_{\mu} ) u - u ( \partial_{\mu} - i l_{\mu} ) u^{\dagger} \right], \\
	U=&\;u^2=\text{exp}\left(i {\phi}/{F_{\pi}}\right) ,
\end{aligned} 
\end{equation}
where $D_{\mu}$ is the covariant derivative introducing external vector fields and axial-vector fields, $u_{\mu}$ is the chiral vielbein and $Q_h = e\, \text{diag} ~ ( 1, 0, 0)$ is the charge operator of the charmed baryon \cite{Guo:2008ns}.

In the numerical analysis, we take $m_c=1.27\,\text{GeV}$, $F_\pi=92.4\,\text{MeV}$, $M_K=494\,\text{MeV}$, $V_{dc}=0.221$, $V_{sc}=0.987$ and the average of the masses for each flavor multiplet, i.e., $m_{\bar{3}}= 2407\,\text{MeV}$, $m_6 = 2535\,\text{MeV}$~\cite{ParticleDataGroup:2022pth}. The mass difference is $\Delta = m_6 - m_{\bar{3}} = 128\,\text{MeV}$. Further,  the values of the various  low energy constants (LECs) $h_{1-3}$ are $h_1 = 0.98$ and $h_2 = -\, 0.60$ 
~\cite{Jiang:2015xqa}. Due to heavy quark spin symmetry, we have $h_3=0$. The magnetic moment couplings, $\alpha_{1-5}$ do not contribute to the EDMs at the order we work here.

\subsection{Construction of the effective $C\!P$-violating Lagrangian}

The effective Lagrangian at the hadron level arising from the dimension-6 terms in Eq.~\eqref{pt_vio_op} is constructed following the same procedure provided in~\cite{Unal:2021lhb}.
The charm-quark-EDM (qEDM) operator, which already contains the electromagnetic field strength tensor $F_{\mu \nu}$, directly induces EDMs of baryons including charm quarks. Only two terms in the leading-order chiral Lagrangian corresponding to EDMs of charm quark baryons contribute.
As in the case of the qEDM operator, there is no light quark content in the charm-quark chromo-EDM (qCEDM) Lagrangian as well. It contains only the heavy charm quark and the gluon field strength tensor $G_{\mu \nu}^{a}$ which makes it invariant under chiral SU(3) transformations.  However, because standard ChPT does not contain a fundamental building block that transforms as a chiral singlet, we have to introduce a new fundamental block $\varphi^+$. To ensure that the chiral singlet $\varphi^+$ violates
$P$ and $T$ one needs to combine $\varphi^+$ with ChPT building blocks to construct $C\!P$-violating terms at the
hadron level.  Further information can be found in e.g. Refs.~\cite{Bsaisou:2014goa, Bsaisou:2014oka,  Unal:2021lhb}.

As for the 4q and 4qLR operators, it can be seen in Eq.~\eqref{pt_vio_op} that there are two types of terms, $\bar{q}q\bar{c}\gamma_5 c+\bar{q}\gamma_5 q\bar{c}c$ and $ \bar{c} \gamma_5 q \bar{q}c+\bar{c} \gamma_5 q \bar{q}c$, including both the heavy charm quark and the light quarks $q=\,u, d, s$. This is different from the qEDM and qCEDM operators. Thus the following matrices can be built up for the transformation properties of $\mathcal{L}_{c,  \mathrm{4q}}^{(6)}$ and $\mathcal{L}_{c,  \mathrm{4qLR}}^{(6)}$ under chiral SU(3) transformations as a new scalar source similar to the quark mass matrix in ordinary ChPT
\begin{equation}
\mathcal{M}^c_1=
\begin{pmatrix}
\kappa_1^{uc} & 0 & 0 \\
0 & \kappa_1^{dc} & 0 \\
0 & 0 & \kappa_1^{sc} 
\end{pmatrix},\quad \quad  	
	\mathcal{N}^c_1=
	\begin{pmatrix}
		0 & 0 & 0 \\
		0 & \rho_1^{dc} V_{dc} & 0 \\
		0 & 0 & \rho_1^{sc} V_{sc}  
	\end{pmatrix}.
	\label{matrices}
\end{equation}
The explicit insertions of the charm quarks allow us to set up the heavy charm baryon matrices $B_{\bar{3}}$ and $B_{6}$ in the effective Lagrangian. One can see that the $\kappa_1$, $\kappa_8$ and $\rho_1$, $\rho_8$  are distinguishable on the quark-level. At the level of chiral EFT, however, they have identical chiral symmetry properties and thus the resulting chiral Lagrangians are identical. Therefore, the effective Lagrangian from the relevant operators combines the effects of these terms. 

The effective Lagrangians resulting from the various $P$- and $T$-violating dimension-6 operators are given by
	\begin{equation}      
	\begin{aligned}                           
	\mathcal{L}_{qEDM}^{\rm eff.}=~& f_1 \langle \bar{B}_{\bar{3}} \sigma^{\mu \nu} \gamma_5 F_{\mu \nu}  B_{\bar{3}} \rangle  +  f_2 \langle \bar{B}_{6} \sigma^{\mu \nu} \gamma_5 F_{\mu \nu}  B_{6} \rangle + \dots\,,\\
\mathcal{L}_{\mathrm{qCEDM}}^{\rm eff.}
%_{q/gCEDM}^{\rm eff}
=~& \varphi^+  \Big [ a_{16} \langle \bar{B}_{\bar{3}} \sigma^{\mu \nu} \gamma_5 F_{\mu \nu}^+ B_{\bar{3}} \rangle +a_{17} \langle \bar{B}_6 \sigma^{\mu \nu} \gamma_5 F_{\mu \nu}^+ B_6 \rangle 
                                             + a_{19} \langle \bar{B}_{\bar{3}}\sigma^{\mu \nu} \gamma_5 B_{\bar{3}} \rangle \langle F_{\mu \nu}^+\rangle \\
                                             & +a_{20} \langle \bar{B}_{6} \sigma^{\mu \nu} \gamma_5 B_6 \rangle \langle F_{\mu \nu}^+ \rangle \Big]  +\dots\,,
     \label{pt_vio_Lagr}                                        
	\end{aligned}                                          
\end{equation} 	

\begin{equation}
	\begin{aligned}                                                                                      
	\mathcal{L}_{\mathrm{4q}}^{\rm eff.}=~& i\kappa_7 \langle \bar{B}_{6} \tilde{\chi}_- B_6 \rangle + i\kappa_8 \langle \bar{B}_6 \tilde{\chi}_- B_{\bar{3}} + h.c.  \rangle 
                                              + \kappa_{11} \langle \bar{B}_{\bar{3}} \tilde{\chi}_+ \sigma^{\mu \nu} \gamma_5 F_{\mu \nu}^+ B_{\bar{3}} \rangle  \\
                                          & +\kappa_{12} \langle \bar{B}_{6} \tilde{\chi}_+ \sigma^{\mu \nu} \gamma_5 F_{\mu \nu}^+ B_{6} \rangle+ \kappa_{14}  \langle \bar{B}_{\bar{3}} \tilde{\chi}_+ \sigma^{\mu \nu} \gamma_5 B_{\bar{3}} \rangle \langle F_{\mu \nu}^+ \rangle  \\
                                          & + \kappa_{15} \langle \bar{B}_{6} \tilde{\chi}_+ \sigma^{\mu \nu} \gamma_5 B_{6} \rangle \langle F_{\mu \nu}^+ \rangle + 
                                           + \kappa_{17} \langle \bar{B}_{\bar{3}} \sigma^{\mu \nu} \gamma_5 F_{\mu \nu}^+ B_{\bar{3}} \rangle \langle \tilde{\chi}_+ \rangle  \\
                                           & + \kappa_{18} \langle \bar{B}_{6} \sigma^{\mu \nu} \gamma_5 F_{\mu \nu}^+ B_{6} \rangle \langle \tilde{\chi}_+ \rangle  
                                          + \kappa_{20} \langle \bar{B}_{\bar{3}} \sigma^{\mu \nu} \gamma_5  B_{\bar{3}} \rangle \langle \tilde{\chi}_+ F_{\mu \nu}^+ \rangle \\ 
                               			  &+ \kappa_{21} \langle \bar{B}_{6} \sigma^{\mu \nu} \gamma_5  B_{6} \rangle \langle \tilde{\chi}_+ F_{\mu \nu}^+ \rangle 
                                           +\dots \, , \\                                     
	\mathcal{L}_{\mathrm{4qLR}}^{\rm eff.}=~& i \rho_2 \langle \bar{B}_6 \hat{\tilde{\chi}}_- B_6 \rangle + i \rho_3 \langle \bar{B}_6 \hat{\tilde{\chi}}_- B_{\bar{3}} + h.c.  \rangle + \rho_{14}  \langle \bar{B}_{\bar{3}} \hat{\tilde{\chi}}_+ \sigma^{\mu \nu} \gamma_5 B_{\bar{3}} \rangle \langle F_{\mu \nu}^+ \rangle \\
                                          & + \rho_{15} \langle \bar{B}_{6} \hat{\tilde{\chi}}_+ \sigma^{\mu \nu} \gamma_5 B_{6} \rangle \langle F_{\mu \nu}^+ \rangle 
                                          + \rho_{17} \langle \bar{B}_{\bar{3}} \sigma^{\mu \nu} \gamma_5 F_{\mu \nu}^+ B_{\bar{3}} \rangle \langle \hat{\tilde{\chi}}_+ \rangle \\ 
                                          & + \rho_{18} \langle \bar{B}_{6} \sigma^{\mu \nu} \gamma_5 F_{\mu \nu}^+ B_{6} \rangle \langle \hat{\tilde{\chi}}_+ \rangle  
                                           +\dots     \,,\label{pt_vio_Lagr}                                        
	\end{aligned}                                          
\end{equation} 
with the definitions
\begin{equation}
	\begin{aligned}
\chi_{\pm}=~& u^{\dagger} \chi u^{\dagger} \pm u \chi ^{\dagger} u ,  \quad \quad \chi= 2 B_0\, \mathrm{diag}(m_u,\,m_d,\,m_s) , \\ 
	\tilde{\chi}_{\pm}= ~& u^{\dagger} \tilde{\chi} u^{\dagger} \pm u \tilde{\chi}^{\dagger} u ,  \quad \quad \tilde{\chi} \equiv   \text{diag} ~ (\kappa^{uc},  \kappa^{dc},  \kappa^{sc})\,,\\ 
	\hat{\tilde{\chi}}_{\pm} = ~& u^{\dagger} \hat{\tilde{\chi}} u^{\dagger} \pm u \hat{\tilde{\chi}}^{\dagger} u ,  \quad \quad \hat{\tilde{\chi}} \equiv   \text{diag} ~ ( 0 ,\text{Re} (V_{dc}) \rho^{dc},\text{Re} (V_{sc}) \rho^{sc}) \,, 
	\label{sources}
	\end{aligned}
\end{equation}
where the ellipsis denotes terms that are either of higher order or irrelevant to the EDM calculations. Here, we give only relevant terms for our calculation. The complete list of operators appearing at the same chiral order can be found in ~\cite{Unal:2021lhb}. In principle, the terms 
\begin{eqnarray}
 \rho_{11} \langle \bar{B}_{\bar{3},v} \hat{\tilde{\chi}}_+ v^{\mu} S^{\nu} F_{\mu \nu}^+ B_{\bar{3},v} \rangle , \nonumber \\
 \rho_{12} \langle \bar{B}_{6,v} \hat{\tilde{\chi}}_+ v^{\mu} S^{\nu} F_{\mu \nu}^+ B_{6,v} \rangle ,  \nonumber \\
 \rho_{20} \langle \bar{B}_{\bar{3},v} v^{\mu} S^{\nu} B_{\bar{3},v} \rangle \langle \hat{\tilde{\chi}}_+ F_{\mu \nu}^+ \rangle , \nonumber \\
  \rho_{21} \langle \bar{B}_{6,v} v^{\mu} S^{\nu} B_{6,v} \rangle \langle \hat{\tilde{\chi}}_+ F_{\mu \nu}^+ \rangle ,  \nonumber
\end{eqnarray}
in 4qLR Lagrangian also contribute to the EDMs, however, because $\langle \hat{\tilde{\chi}}_+ Q_h \rangle$ results in zero, we do not display those terms in Eq.~\eqref{pt_vio_Lagr}. Here, one should notice that the constants $\kappa^{uc, dc, sc}$ and $\rho^{dc, sc}$ get hold of both the color-singlet and color-octet terms whose chiral Lagrangians are identical. 

Making use of the heavy-baryon (HB) formulation of baryon ChPT describing systems with a single heavy quark~\cite{Jenkins:1990jv, Bernard:1992qa} is convenient for the calculation of EDMs. The $C\!P$-violating Lagrangian in the HB limit
\begin{equation}
	\begin{aligned}                     
	\mathcal{L}_{\text{free}}^{(1)}= ~&\frac{1}{2}\langle \bar{B}_{\bar{3},v}( i v\cdot D) B_{\bar{3},v} \rangle +\langle \bar{B}_{6,v}(i v \cdot D-\Delta) B_{6,v} \rangle,  \\                                                                                                                
	\mathcal{L}_{\text{int}}=~ ~ &h_1 \langle \bar{B}_{6,v} u_{\mu} S^{\mu} {B}_{6,v} \rangle+ h_2\langle \bar{B}_{6,v}u_{\mu} S^{\mu} {B}_{\bar{3},v}+h.c. \rangle , \\                                                                                                                                                    
	\mathcal{L}_{B \gamma}^{(2)}=~&2 \varepsilon^{\mu  \nu \rho \sigma} \Big[\alpha_1 \langle \bar{B}_{\bar{3},v} v_{\rho} S_{\sigma} F_{\mu \nu}^+ B_{\bar{3},v} \rangle + \alpha_2 \langle \bar{B}_{6,v} v_{\rho} S_{\sigma} F_{\mu \nu}^+ B_{6,v} \rangle + \alpha_3 \langle \bar{B}_{6,v} v_{\rho} S_{\sigma} F_{\mu \nu}^+ B_{\bar{3},v} + h.c.  \rangle \\
                                               & + \alpha_4 \langle \bar{B}_{\bar{3},v} v_{\rho} S_{\sigma}B_{\bar{3},v} \rangle \langle F_{\mu \nu}^+ \rangle + \alpha_5 \langle \bar{B}_{6,v} v_{\rho} S_{\sigma} B_{6,v} \rangle \langle F_{\mu \nu}^+ \rangle \Big],
    \label{heavy Lagr}                                           
	\end{aligned}
\end{equation}

\begin{equation}
	\begin{aligned}              
	\mathcal{L}_{\mathrm{qEDM}}^{\rm eff.}=~& 4 i \Big[ f_1 \langle \bar{B}_{\bar{3},v} v^{\mu} S^{\nu} F_{\mu \nu} B_{\bar{3},v} \rangle  + f_2 \langle \bar{B}_{6,v} v^{\mu} S^{\nu} F_{\mu \nu} B_{6,v} \rangle \Big], \\         
		\mathcal{L}_{\mathrm{qCEDM}}^{\rm eff.}
	%_{q/gCEDM}^{\rm eff}
	=~& 4 i \varphi^+ \Big [ a_{16} \langle \bar{B}_{\bar{3},v} v^{\mu} S^{\nu} F_{\mu \nu}^+ B_{\bar{3},v} \rangle +a_{17} \langle \bar{B}_{6,v} v^{\mu} S^{\nu}  F_{\mu \nu}^+ B_{6,v} \rangle + 
                                           + a_{19} \langle \bar{B}_{\bar{3},v}v^{\mu} S^{\nu}  B_{\bar{3},v} \rangle \langle F_{\mu \nu}^+\rangle \\
                                           & + a_{20} \langle \bar{B}_{6,v} v^{\mu} S^{\nu}  B_{6,v} \rangle \langle F_{\mu \nu}^+ \rangle \Big] + \dots \, , \\                                               
	\mathcal{L}_{\mathrm{4q}}^{\rm eff.}=~& i\kappa_7 \langle \bar{B}_{6,v} \tilde{\chi}_- B_{6,v} \rangle + i\kappa_8 \langle \bar{B}_{6,v} \tilde{\chi}_- B_{\bar{3},v} + h.c.  \rangle + 4 i \Big[ \kappa_{11} \langle \bar{B}_{\bar{3},v} \tilde{\chi}_+ v^{\mu} S^{\nu} F_{\mu \nu}^+ B_{\bar{3},v} \rangle \\
	                                            &+\kappa_{12} \langle \bar{B}_{6,v} \tilde{\chi}_+ v^{\mu} S^{\nu} F_{\mu \nu}^+ B_{6,v} \rangle + \kappa_{14}  \langle \bar{B}_{\bar{3},v} \tilde{\chi}_+ v^{\mu} S^{\nu} B_{\bar{3},v} \rangle \langle F_{\mu \nu}^+ \rangle  \\
                                                & + \kappa_{15} \langle \bar{B}_{6,v} \tilde{\chi}_+ v^{\mu} S^{\nu} B_{6,v} \rangle \langle F_{\mu \nu}^+ \rangle + \kappa_{17} \langle \bar{B}_{\bar{3},v} v^{\mu} S^{\nu} F_{\mu \nu}^+ B_{\bar{3},v} \rangle \langle \tilde{\chi}_+ \rangle \\
                                                & + \kappa_{18} \langle \bar{B}_{6,v} v^{\mu} S^{\nu} F_{\mu \nu}^+ B_{6,v} \rangle \langle \tilde{\chi}_+ \rangle + \kappa_{20} \langle \bar{B}_{\bar{3},v} v^{\mu} S^{\nu} B_{\bar{3},v} \rangle \langle \tilde{\chi}_+ F_{\mu \nu}^+ \rangle \\
                                                & + \kappa_{21} \langle \bar{B}_{6,v} v^{\mu} S^{\nu} B_{6,v} \rangle \langle \tilde{\chi}_+ F_{\mu \nu}^+ \rangle \Big] + \dots \, , \\
	\mathcal{L}_{\mathrm{4qLR}}^{\rm eff.}=~& i \rho_2 \langle \bar{B}_{6,v} \hat{\tilde{\chi}}_- B_{6,v} \rangle + i \rho_3 \langle \bar{B}_{6,v} \hat{\tilde{\chi}}_- B_{\bar{3},v} + h.c.  \rangle 
                                         	 + 4 i \Big[ \rho_{14}  \langle \bar{B}_{\bar{3},v} \hat{\tilde{\chi}}_+ v^{\mu} S^{\nu} B_{\bar{3},v} \rangle \langle F_{\mu \nu}^+ \rangle  \\
                                         	  & + \rho_{15} \langle \bar{B}_{6,v} \hat{\tilde{\chi}}_+ v^{\mu} S^{\nu} B_{6,v} \rangle \langle F_{\mu \nu}^+ \rangle + \rho_{17} \langle \bar{B}_{\bar{3},v} v^{\mu} S^{\nu} F_{\mu \nu}^+ B_{\bar{3},v} \rangle \langle \hat{\tilde{\chi}}_+ \rangle  \\
                                         	  & + \rho_{18} \langle \bar{B}_{6,v} v^{\mu} S^{\nu} F_{\mu \nu}^+ B_{6,v} \rangle \langle \hat{\tilde{\chi}}_+ \rangle \Big]+ \dots \,  ,                          
	\end{aligned}                                          
\end{equation}
with the four velocity $v^{\mu}$, the Pauli-Lubanski spin operator $S^{\mu}=-\gamma_5(\gamma^{\mu} \slashed{v}-v^{\mu})/2$ and the mass difference $\Delta=m_6-m_{\bar{3}}$. 
Only terms linear in the Goldstone bosons are sufficient at the order we work on. Higher-order terms containing more Goldstone bosons are embedded in the ellipses. The chiral singlet $\varphi^+$ can be absorbed into the $a_i$ LECs, as it can only contribute as an overall constant. 

\begin{figure}[t]
	\centering
	\includegraphics[width=0.8\textwidth]{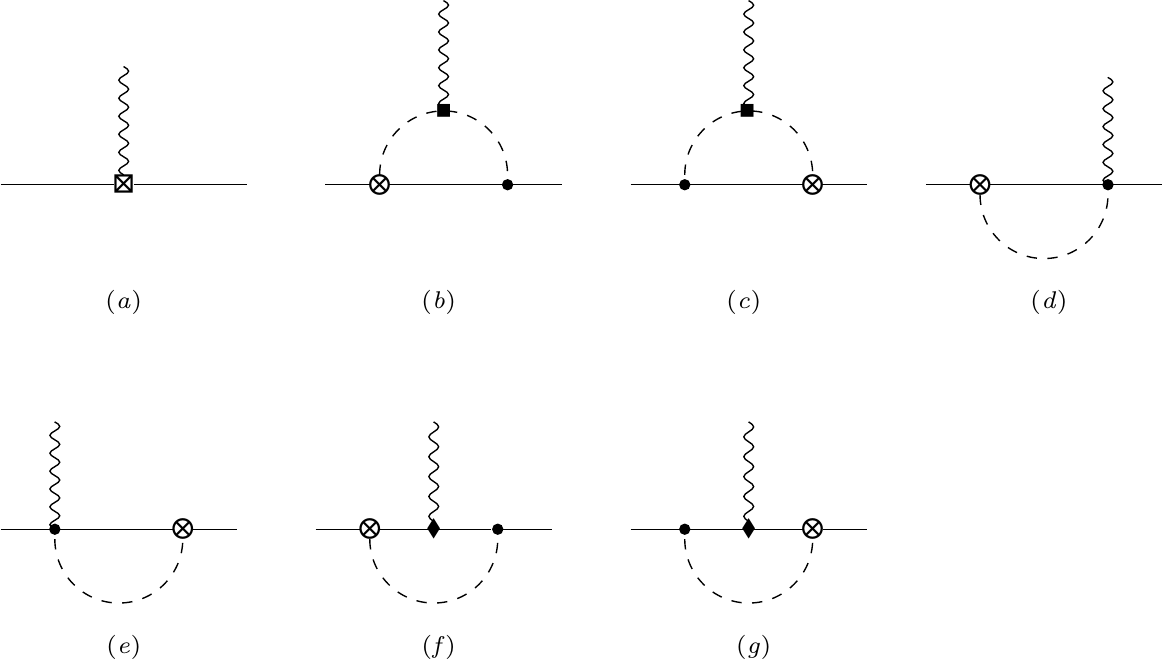}
	\caption{Diagrams that contribute to the EDMs of the spin-$1/2$ anti-triplet and sextet
          $c$-baryons after renormalization.  The solid, dashed, and wavy lines depict the baryons, mesons, and photons, respectively.
	  The filled circles, diamonds, and squares are the first-order vertices generated by the meson-baryon Lagrangian and second-order mesonic vertices, in order. $C\!P$-violating vertices at $\mathcal{O}(\delta^0)$
          and $\mathcal{O}(\delta^2)$ are represented by $\otimes$ and $\boxtimes$, respectively. }
	\label{fig:diag}
\end{figure}

Figure \ref{fig:diag} demonstrates tree level and one-loop Feynman diagrams that generate a non-vanishing contribution to the $P$- and $T$-violating form factor
of the charm baryons up to the order $\mathcal{O}(\delta^2)$, where $\delta$ is a generic small mass or momentum.
We apply the modified minimal subtraction scheme of HBChPT ($\widetilde{\text{MS}}$)~\cite{tHooft:1973mfk, Weinberg:1973xwm,Gasser:1983yg}
and calculate the loop diagrams in the framework of dimensional
regularization at the renormalization scale $\lambda=1$~GeV. 
Tree-level $C\!P$-odd diagrams at order $\mathcal{O}(\delta^2)$
denoted in diagram (a) receive contributions from all the $C\!P$-violating operators. The diagrams $(b)$-$(g)$ in Figure~\ref{fig:diag} show one-loop diagrams at leading $\mathcal{O}(\delta^2)$.
We apply the standard power counting to the renormalized diagrams (see, e.g.,~Ref.\cite{Scherer:2012zzd}). In other words, an interaction vertex obtained from an $\mathcal{O}(\delta^n$) Lagrangian is counted as order $\delta^n$, a meson propagator as order $\delta^{-2}$, a baryon propagator as order $\delta^{-1}$, and the integration of a loop as order $\delta^4$. For the $C\!P$-odd vertices, the chiral order of the sources is counted as $\mathcal{O}(\delta^0)$.

\section{The $P$- and $T$-violating form factor}\label{sec:loop}

In the heavy baryon approach and in the Breit frame the $P$- and $T$-violating form factor $D_{B_h}^{\gamma}(q^2)$ is described through
%--
\begin{eqnarray}
 \Braket{B_h (p_{f})|J_{\text{EDM}, \nu}|B_{h} (p_i)} = -2 i D_{B_h}^{\gamma}(q^2) \bar{B}_{v} v_{\nu} (S \cdot q) B_{v} 
\end{eqnarray}	
%--
where $q = p_f-p_i$ denotes the four-momentum transfer (see e.g., ~\cite{Borasoy:2000pq}). The EDM is thus defined by 
\begin{eqnarray}
d_{B_h}^\gamma = D_{B_h}^{\gamma}(q^2=0).
\end{eqnarray}	
The four-velocity is set $v_{\mu}=(1,\textbf{0})$ and $v\cdot p_i=v\cdot p_f$ in the Breit frame.
The contributions from the contact interactions in Figure~\ref{fig:diag}-(a) are presented in Tables \ref{tab:tree1}-\ref{tab:tree3}.
%--
 \begin{table}[t]
 	\centering
	\caption{Tree-level contributions from the qEDM and qCEDM operators of the charm baryons.}
	\begin{tabular}{ccc}
		\hline
{Baryons}                   &  {qEDM}          & {qCEDM}     \\
		\hline  
{$\Lambda_{c}^+$}  
  & {$ 4 f_1$} 
  &  {$ 4 e (a_{16} + 2 a_{19}) $} \\
{$\Xi_{c}^0$}           
  & {$ 4 f_1$} 
  &  {$ 8e a_{19}$} \\
 {$\Xi_{c}^+$}          
  & {$ 4 f_1$} 
  & {$4 e (a_{16} + 2 a_{19})$} \\
		\hline
 {$\Sigma_{c}^{0}$}
  & {$ 2 f_2$} 
  &  {$ 4 e  a_{20} $} \\
{$\Sigma_{c}^{+}$}       
  & {$ 2 f_2$} 
  &  {$ e (a_{17} + 4 a_{20}) $} \\
{$\Sigma_{c}^{++}$}       
  & {$ 2 f_2$} 
  &  {$ 2e  (a_{17}+2 a_{20})$} \\
{$\Xi_{c}^{'0}$}              
  & {$ 2 f_2$} 
  &  {$ 4 e  a_{20}$} \\
{$\Xi_{c}^{'+}$}              
  & {$ 2 f_2$} 
  &  {$e (a_{17} + 4 a_{20}) $} \\
{$\Omega_{c}^{0}$}      
  & {$ 2 f_2 $} 
  &  {$ 4 e a_{20} $} \\
		\hline
	\end{tabular}
	\label{tab:tree1}
\end{table}
%--

As in the case of the $b$-baryons~\cite{Unal:2021lhb}, the EDMs of the $c$-baryons also receive contributions from the Goldstone bosons. For the qEDM and qCEDM operators, we observe that the meson-loops appear at a higher order and only the tree-level diagrams provide the contributions at the order we work. The loop contributions at the second chiral order are obtained from the 4q- and 4qLR operators and the LECs of those tree-level contributions absorb the relevant loop divergences. 

Most of the diagrams depicted in Figure~\ref{fig:diag} are either proportional to $S \cdot v =0$, or $v\cdot q=0$. Furthermore, some of them mutually cancel each other. After an explicit calculation in HBChPT,  it has been seen that only the diagrams $(b)$ and $(c)$ are non-vanishing and yield the following contributions
\begin{equation}
	\begin{aligned} 
	D^{\gamma}_{B_h, (b)}(q^2)= ~&\frac{A_{b_i}}{2} \int_0^1 dx \, \frac{x}{\tilde{M}_i}\frac{\partial}{\partial \tilde{M}_i}J_1(\tilde{w}, \tilde{M}_i), \\
	D^{\gamma}_{B_h, (c)}(q^2)=  ~&\frac{A_{c_i}}{2} \int_0^1 dx \, \frac{x-1}{\tilde{M}_i}\frac{\partial}{\partial \tilde{M}_i}J_1(\tilde{w}, \tilde{M}_i),  \quad i = 1, 2, 3, 4, 
     \label{res4q-edm}
	\end{aligned}
\end{equation} 
where $\tilde{w}=-\Delta$, for a sextet particle, $\tilde{w}=0$ for an anti-triplet particle inside the loop, $\tilde{M}_i(x)=\sqrt{x(x-1)q^2+M_i^2}$, with $M_i$ being charged meson masses and $J_1$ is the loop function (see Appendix~\ref{sec:AppendixC}). As many of the $A_{b_i}$ and $A_{c_i}$ coefficients are identical to each other except for their sign. The remaining terms including these coefficients are given in Appendix~\ref{sec:AppendixA}. 
 \begin{table}[t]
 	\centering
	\caption{Tree-level contributions from the 4q operator of the charm baryons. } 
\begin{tabular}{cc}
		\hline
{Baryons}             &          {4q} \\
		\hline  
{$\Lambda_{c}^+$}  
  & {$ 8e[ (\kappa_{11}+ 2 \kappa_{20}) \kappa^{uc} + \kappa_{14}(\kappa^{uc}+\kappa^{dc}) + \kappa_{17} (\kappa^{uc}+\kappa^{dc}+\kappa^{sc})] $} 
  \\
{$\Xi_{c}^0$}           
  & {$ 8e[\kappa_{14}(\kappa^{dc}+\kappa^{sc}) + 2 \kappa_{20} \kappa^{uc}] $}  \\
  
{$\Xi_{c}^+$}          
  & {$ 8e[ (\kappa_{11}+ 2 \kappa_{20}) \kappa^{uc} + \kappa_{14}(\kappa^{uc}+\kappa^{sc}) + \kappa_{17} (\kappa^{uc}+\kappa^{dc}+\kappa^{sc})] $} \\
  
		\hline
{$\Sigma_{c}^{0}$}
  & {$ 8e (\kappa_{15} \kappa^{dc}+\kappa_{21} \kappa^{uc}) $} \\
  
{$\Sigma_{c}^{+}$}       
  & {$ 4e[ (\kappa_{12}+ 2 \kappa_{21}) \kappa^{uc} + \kappa_{15}(\kappa^{uc}+\kappa^{dc}) + \kappa_{18} (\kappa^{uc}+\kappa^{dc}+\kappa^{sc})] $} \\
  
{$\Sigma_{c}^{++}$}       
  & {$8e[ (\kappa_{12}+ \kappa_{15}+ \kappa_{21}) \kappa^{uc} + \kappa_{18}(\kappa^{uc}+\kappa^{dc}+\kappa^{sc})] $} \\
  
{$\Xi_{c}^{'0}$}              
  & {$ 4e[\kappa_{15}(\kappa^{dc}+\kappa^{sc}) + 2\kappa_{21} \kappa^{uc}] $} \\
  
{$\Xi_{c}^{'+}$}              
  & {$ 4e[ (\kappa_{12}+ 2 \kappa_{21}) \kappa^{uc} + \kappa_{15}(\kappa^{uc}+\kappa^{sc}) + \kappa_{18} (\kappa^{uc}+\kappa^{dc}+\kappa^{sc})] $}  \\
 
{$\Omega_{c}^{0}$}      
  & {$ 8e (\kappa_{15} \kappa^{sc}+\kappa_{21} \kappa^{uc}) $} \\

		\hline
	\end{tabular}
	\label{tab:tree2}
\end{table}
%--
 \begin{table}[t]
	\centering
	\caption{Tree-level contributions from the 4qLR operator of the charm baryons. } 
	\begin{tabular}{cc}
		\hline
		{Baryons}        &       {4qLR}   \\
		\hline  
		{$\Lambda_{c}^+$}  
		
		& {$ 8e [\text{Re}(V_{dc})(\rho_{14} +\rho_{17})  \rho^{dc} + \text{Re}(V_{sc}) \rho_{17} \rho^{sc} ] $} \\
		{$\Xi_{c}^0$}           
		
		& {$ 8e [\text{Re}(V_{dc})\rho^{dc} + \text{Re}(V_{sc})\rho^{sc} ] \rho_{14} $}  \\
		{$\Xi_{c}^+$}          
		
		& {$8e [\text{Re}(V_{sc})(\rho_{14} +\rho_{17})  \rho^{sc} + \text{Re}(V_{dc}) \rho_{17} \rho^{dc} ] $} \\
		\hline
		{$\Sigma_{c}^{0}$}
		& {$ 8e \text{Re}(V_{dc})\rho_{15} \rho^{dc} $} \\
		{$\Sigma_{c}^{+}$}       

		& {$ 4e [\text{Re}(V_{dc})(\rho_{15} +\rho_{18})  \rho^{dc} + \text{Re}(V_{sc}) \rho_{18} \rho^{sc} ] $} \\
		{$\Sigma_{c}^{++}$}       

		& {$ 8e [\text{Re}(V_{dc})\rho^{dc} + \text{Re}(V_{sc})\rho^{sc} ] \rho_{18} $} \\
		{$\Xi_{c}^{'0}$}              

		& {$ 4e [\text{Re}(V_{dc})\rho^{dc} + \text{Re}(V_{sc})\rho^{sc} ] \rho_{15} $} \\
		{$\Xi_{c}^{'+}$}              

		& {$ 4e [\text{Re}(V_{sc})(\rho_{15} +\rho_{18})  \rho^{sc} + \text{Re}(V_{dc}) \rho_{18} \rho^{dc}] $} \\
		{$\Omega_{c}^{0}$}      

		& {$ 8e \text{Re}(V_{sc})\rho_{15} \rho^{sc} $} \\
		\hline
	\end{tabular}
	\label{tab:tree3}
\end{table}
%--
The results for the 4q operator are

\begin{equation}
\vspace{150pt}
	\begin{aligned} 
		d^{\gamma}_{\Lambda_c^+, 4q}= ~& 8e\Big [(\kappa_{11}+ 2 \kappa_{20}) \kappa^{uc} + \kappa_{14}(\kappa^{uc}+\kappa^{dc}) + \kappa_{17} (\kappa^{uc}+\kappa^{dc}+\kappa^{sc})\Big] \\
		  &+ \frac{e g_2 \kappa_8 ( \kappa^{uc}+\kappa^{sc})}{32 \pi^2 F_{\pi}^2} F_{M_K}^{(2)} ,
          \\                                  
		d^{\gamma}_{\Xi_c^0,\mathrm{4q}}= ~& 8e\Big [\kappa_{14}(\kappa^{dc}+\kappa^{sc}) + 2 \kappa_{20} \kappa^{uc}] - \frac{e g_2  \kappa_8 ( \kappa^{uc}+\kappa^{dc})}{32 \pi^2 F_{\pi}^2}  F_{M_{\pi}}^{(2)} \\
		&- \frac{e g_2  \kappa_8 ( \kappa^{uc}+\kappa^{sc})}{32 \pi^2 F_{\pi}^2}  F_{M_K}^{(2)} , \\
		    d^{\gamma}_{\Xi_c^+,\mathrm{4q}}= ~& 8e \Big [ (\kappa_{11}+ 2 \kappa_{20}) \kappa^{uc} + \kappa_{14}(\kappa^{uc}+\kappa^{sc}) + \kappa_{17} (\kappa^{uc}+\kappa^{dc}+\kappa^{sc}) \Big] \\
		& + \frac{e g_2  \kappa_8 (\kappa^{uc}+\kappa^{dc})} {32 \pi^2 F_{\pi}^2} F_{M_{\pi}}^{(2)} , 
		 \\   
     d^{\gamma}_{\Sigma_c^0,\mathrm{4q}}= ~& 8e (\kappa_{15} \kappa^{dc}+\kappa_{21} \kappa^{uc}) - \frac{e g_1  \kappa_7 (\kappa^{uc}+\kappa^{dc})}{32 \pi^2 F_{\pi}^2}  F_{M_{\pi}}^{(2)} 
          - \frac{e g_2 \kappa_8 (\kappa^{uc}+\kappa^{dc}) }{16 \pi^2 F_{\pi}^2} F_{M_{\pi}}^{(1)}  ,
		\\                                  		 		   	     
	d^{\gamma}_{\Sigma_c^+,\mathrm{4q}}=  ~& 4e \Big [ (\kappa_{12}+ 2 \kappa_{21}) \kappa^{uc} + \kappa_{15}(\kappa^{uc}+\kappa^{dc}) + \kappa_{18} (\kappa^{uc}+\kappa^{dc}+\kappa^{sc}) \Big] \\
		   &  + \frac{e g_1 \kappa_7 ( \kappa^{uc}+\kappa^{sc})}{64 \pi^2 F_{\pi}^2}  F_{M_K}^{(2)}   
	         + \frac{e g_2 \kappa_8 ( \kappa^{uc}+\kappa^{sc})}{32 \pi^2 F_{\pi}^2}  F_{M_K}^{(1)} ,
	      \\
	      d^{\gamma}_{\Sigma_c^{++},\mathrm{4q}}=  ~& 8e \Big[ (\kappa_{12}+ \kappa_{15}+ \kappa_{21}) \kappa^{uc} + \kappa_{18}(\kappa^{uc}+\kappa^{dc}+\kappa^{sc}) \Big] 
		 + \frac{e g_1 \kappa_7 ( \kappa^{uc}+\kappa^{dc})}{32 \pi^2 F_{\pi}^2} 
		 F_{M_\pi}^{(2)}  \\
		 & + \frac{e g_1 \kappa_7 ( \kappa^{uc}+\kappa^{sc}) }{32 \pi^2 F_{\pi}^2}  F_{M_K}^{(2)}                                           
           + \frac{e g_2 \kappa_8 ( \kappa^{uc}+\kappa^{dc})}{16 \pi^2 F_{\pi}^2} 
		 F_{M_\pi}^{(1)}  
		 + \frac{e g_2 \kappa_8 ( \kappa^{uc}+\kappa^{sc}) }{16 \pi^2 F_{\pi}^2}  F_{M_K}^{(1)} ,    \\
d^{\gamma}_{\Xi_c^{'0},\mathrm{4q}}=  ~&  4e \Big [\kappa_{15}(\kappa^{dc}+\kappa^{sc}) + 2\kappa_{21} \kappa^{uc}] \Big] 
		    - \frac{e g_1 \kappa_7 ( \kappa^{uc}+\kappa^{dc})}{64 \pi^2 F_{\pi}^2} 
		 F_{M_\pi}^{(2)}  \\
		 & - \frac{e g_1 \kappa_7 ( \kappa^{uc}+\kappa^{sc}) }{64 \pi^2 F_{\pi}^2}  F_{M_K}^{(2)}                               
           - \frac{e g_2 \kappa_8 ( \kappa^{uc}+\kappa^{dc})}{32 \pi^2 F_{\pi}^2} 
		 F_{M_\pi}^{(1)}  
		 - \frac{e g_2 \kappa_8 ( \kappa^{uc}+\kappa^{sc}) }{32 \pi^2 F_{\pi}^2}  F_{M_K}^{(1)} ,                                  \\                             
d^{\gamma}_{\Xi_c^{'+},\mathrm{4q}}=  ~&  4e \Big[ (\kappa_{12}+ 2 \kappa_{21}) \kappa^{uc} + \kappa_{15}(\kappa^{uc}+\kappa^{sc}) + \kappa_{18} (\kappa^{uc}+\kappa^{dc}+\kappa^{sc}) \Big]
		\\
		& + \frac{e g_1 \kappa_7 (\kappa^{uc}+\kappa^{dc})} {64 \pi^2 F_{\pi}^2}  F_{M_{\pi}}^{(2)} + \frac{e g_2 \kappa_8 (\kappa^{uc}+\kappa^{dc})} {32 \pi^2 F_{\pi}^2} F_{M_{\pi}}^{(1)} ,
		\\	
	d^{\gamma}_{\Omega_c^{0},\mathrm{4q}}=  ~& 8e (\kappa_{15} \kappa^{sc}+\kappa_{21} \kappa^{uc}) 
		 - \frac{e g_1 \kappa_7 ( \kappa^{uc}+\kappa^{sc}) }{32 \pi^2 F_{\pi}^2}  F_{M_K}^{(2)} 
		 - \frac{e g_2 \kappa_8 ( \kappa^{uc}+\kappa^{sc})}{16 \pi^2 F_{\pi}^2}  F_{M_K}^{(1)} \, .
		\label{all-edm-4q}
	\end{aligned}
\end{equation} 
%--
For the 4qRL operator, the results are
%--
\begin{equation}
	\begin{aligned} 
		d^{\gamma}_{\Lambda_c^+, 4qLR}= ~& 8e [\text{Re}(V_{dc})(\rho_{14} +\rho_{17})  \rho^{dc} + \text{Re}(V_{sc}) \rho_{17} \rho^{sc} ]  
			 + \frac{e \text{Re}(V_{sc}) g_2  \rho_3 \rho^{sc}}{32 \pi^2 F_{\pi}^2} F_{M_K}^{(2)} ,
			 \\
		d^{\gamma}_{\Xi_c^0,\mathrm{4qLR}}= ~& 8e[\text{Re}(V_{dc})\rho^{dc} + \text{Re}(V_{sc})\rho^{sc} ] \rho_{14}
			  - \frac{e \text{Re}(V_{sc})g_2 \rho_3 \rho^{sc}} {32 \pi^2 F_{\pi}^2} F_{M_K}^{(2)}   \\ 
		      & - \frac{e \text{Re}(V_{dc})g_2 \rho_3 \rho^{dc}} {32 \pi^2 F_{\pi}^2} F_{M_\pi}^{(2)} ,  
		     \\
		    d^{\gamma}_{\Xi_c^+,\mathrm{4qLR}}= ~&8e [\text{Re}(V_{sc})(\rho_{14} +\rho_{17})  \rho^{sc} + \text{Re}(V_{dc}) \rho_{17} \rho^{dc} ]
		 + \frac{e \text{Re}(V_{dc})g_2 \rho_3 \rho^{dc}} {32 \pi^2 F_{\pi}^2} F_{M_\pi}^{(2)},  
		    \\   
       d^{\gamma}_{\Sigma_c^0,\mathrm{4qLR}}= ~& 8e \text{Re}(V_{dc})\rho_{15} \rho^{dc}  
		- \frac{e \text{Re}(V_{dc}) g_1 \rho_2 \rho^{dc}}{32 \pi^2 F_{\pi}^2} F_{M_\pi}^{(2)} 
		- \frac{e \text{Re}(V_{dc}) g_2 \rho_3 \rho^{dc}}{16 \pi^2 F_{\pi}^2} F_{M_\pi}^{(1)}  ,
		\\                                  		 		   	     
	d^{\gamma}_{\Sigma_c^+,\mathrm{4qLR}}=  ~& 4e [\text{Re}(V_{dc})(\rho_{15} +\rho_{18})  \rho^{dc} + \text{Re}(V_{sc}) \rho_{18} \rho^{sc} ] 
			 + \frac{e \text{Re}(V_{sc})g_1 \rho_2 \rho^{sc}}{64 \pi^2 F_{\pi}^2}F_{M_K}^{(2)} \\
	        & + \frac{e \text{Re}(V_{sc})g_2 \rho_3 \rho^{sc}}{32 \pi^2 F_{\pi}^2} F_{M_K}^{(1)} ,
	     \\
	      d^{\gamma}_{\Sigma_c^{++},\mathrm{4qLR}}=  ~& 8e [\text{Re}(V_{dc})\rho^{dc} + \text{Re}(V_{sc})\rho^{sc} ]   \rho_{18}
		 + \frac{e \text{Re}(V_{dc}) g_1 \rho_2 \rho^{dc}}{32 \pi^2 F_{\pi}^2} 
		 F_{M_\pi}^{(2)}  \\
		 & + \frac{e \text{Re}(V_{sc}) g_1 \rho_2 \rho^{sc}}{32 \pi^2 F_{\pi}^2} F_{M_K}^{(2)}                         
             + \frac{e \text{Re}(V_{dc}) g_2 \rho_3 \rho^{dc}}{16 \pi^2 F_{\pi}^2} 
		 F_{M_\pi}^{(1)}  
		 + \frac{e \text{Re}(V_{sc}) g_2 \rho_3 \rho^{sc}}{16 \pi^2 F_{\pi}^2}   F_{M_K}^{(1)} ,                          
          \\                             
		d^{\gamma}_{\Xi_c^{'0},\mathrm{4qLR}}=  ~&  4e [\text{Re}(V_{dc})\rho^{dc} + \text{Re}(V_{sc})\rho^{sc} ] \rho_{15} 
			 - \frac{e \text{Re}(V_{dc}) g_1 \rho_2 \rho^{dc}}{64 \pi^2 F_{\pi}^2} 
		 F_{M_\pi}^{(2)}  \\
		 & - \frac{e \text{Re}(V_{sc}) g_1 \rho_2 \rho^{sc}}{64 \pi^2 F_{\pi}^2} F_{M_K}^{(2)}                         
              -  \frac{e \text{Re}(V_{dc}) g_2 \rho_3 \rho^{dc}}{32 \pi^2 F_{\pi}^2} 
		 F_{M_\pi}^{(1)}  
		 - \frac{e \text{Re}(V_{sc}) g_2 \rho_3 \rho^{sc}}{32 \pi^2 F_{\pi}^2}   F_{M_K}^{(1)} ,                          
          \\                     
		d^{\gamma}_{\Xi_c^{'+},\mathrm{4qLR}}=  ~& 4e [\text{Re}(V_{sc})(\rho_{15} +\rho_{18})  \rho^{sc} + \text{Re}(V_{dc}) \rho_{18} \rho^{dc} ]
		 + \frac{e \text{Re}(V_{dc}) g_1 \rho_2 \rho^{dc}}{64 \pi^2 F_{\pi}^2} F_{M_{\pi}}^{(2)}  
	     \\
		 & + \frac{e \text{Re}(V_{dc}) g_2 \rho_3 \rho^{dc}}{32 \pi^2 F_{\pi}^2} F_{M_\pi}^{(1)} ,
		\\	
	d^{\gamma}_{\Omega_c^{0},\mathrm{4qLR}}=  ~&8e \text{Re}(V_{sc})\rho_{15} \rho^{sc} 
		 - \frac{e \text{Re}(V_{sc}) g_1 \rho_2 \rho^{sc}}{32 \pi^2 F_{\pi}^2} F_{M_K}^{(2)} 
		 - \frac{e \text{Re}(V_{sc}) g_2 \rho_3 \rho^{sc}}{16 \pi^2 F_{\pi}^2} F_{M_K}^{(1)} \, ,
		\label{all-edm-4qLR}
	\end{aligned}
\end{equation} 
where the loop functions are defined as 
%--
\begin{equation}
      \begin{aligned}
F_{M_\pi}^{(1)} = ~& 1 + 32 \pi^2 L + 2 \text{ln}\Bigg[\frac{M_\pi}{\lambda} \Bigg] ,  \\
F_{M_\pi}^{(2)} = ~& 1 + 32 \pi^2 L + 2 \text{ln}\Bigg[\frac{M_{\pi}}{\lambda}\Bigg] + \frac{2 \Delta}{\sqrt{\Delta^2 - M_{\pi}^2 }} \text{ln}\Bigg[ \frac{\Delta}{M_{\pi}}  + \sqrt{\frac{\Delta^2}{M_{\pi}^2} -1 } \Bigg] , \\
F_{M_K}^{(1)} = ~& 1 + 32 \pi^2 L + 2 \text{ln}\Bigg[\frac{M_K}{\lambda} \Bigg] , \\
F_{M_K}^{(2)} = ~& 1 + 32 \pi^2 L + 2 \text{ln}\Bigg[\frac{M_K}{\lambda}\Bigg] + \frac{2 \Delta \text{Arccos}\big[\frac{\Delta}{M_K}\big]}{\sqrt{M_K^2-\Delta^2}}\,.
     \end{aligned}
\end{equation}

\section{Patterns and sizes of charm baryon EDMs}
Some patterns, which provide to see the relative sizes of the various EDMs, can also be obtained for the charm baryons. For the charm-quark qEDM at $\mathcal{O}({\delta^2})$, the EDMs of all baryons in the triplet and the sextet are described only by $f_1$ or $f_2$. For the qCEDM, $d_{\Xi_c^0}$ is different from $d^\gamma_{\Lambda_c^+}=d^\gamma_{\Xi_c^+}$ in the triplet. On the other hand, $d^\gamma_{\Sigma_c^0} = d^\gamma_{\Xi_c^{'0}} =  d^\gamma_{\Omega_c^0}$,  and $d^\gamma_{\Sigma_c^+} = d^\gamma_{\Xi_c^{'+}}$ are obtained in the sextet. It appears that the neutral states differ from the charged states due to an insertion of the quark charge matrix which generates the EDM for the qCEDM. While the EDMs of singly charm baryons with different charges differ for the qCEDM, this is not valid for the qEDM, because it has already photons.

For the 4qLR operator, while the tree-level contributions to the triplet and sextet EDMs display identical patterns as that of the qCEDM, the loop contributions trigger differences. In the triplet, loop contributions to the EDMs include two different CKM matrix elements and different Goldstone bosons, resulting in the EDMs of the charged baryons to differ. We obtain the difference of the charged states in the triplet is nonzero and finite
\begin{equation}
\begin{aligned}
d^\gamma_{\Lambda_c^+,\mathrm{4qLR}} - d^\gamma_{\Xi_c^+,\mathrm{4qLR}} = 
\, & 8e \rho_{14} \Big[ \mathrm{Re}(V_{dc}) \rho^{dc} - \mathrm{Re}(V_{sc}) \rho^{sc}\Big] \\
&
- \frac{e \rho_2 \rho_3 }{32 \pi^2 F_\pi^2} \Bigg( \mathrm{Re}(V_{dc}) \rho^{dc} F_{M_{\pi}}^{(2)} -\mathrm{Re}(V_{sc}) \rho^{sc} F_{M_{K}}^{(2)} \Bigg) \, ,
\end{aligned}
\end{equation}
which disappears considering only qCEDM. Degeneracies that are seen in the qCEDM for the charged and the neutral sextet baryons are not valid anymore because of the difference in the flavor structures of the underlying operator.  

As with the $b$-baryons EDM calculation, the 4q operators exhibit a different pattern of EDMs, which are strictly dependent on their flavor structure. Here, we see from Eq.~(\ref{matrices}) that the chiral symmetry properties of $\kappa^{dc}$ and $\kappa^{sc}$ are identical to the 4qLR operator $\rho^{dc}$ and $\rho^{sc}$. Therefore,  for the $\kappa^{dc}$ and $\kappa^{sc}$ sources, the same pattern of EDMs appears as for the 4qLR as well. 

%--
To determine the absolute sizes of the EDMs of charm baryons, we need to estimate the various LECs in EDM expressions. In order to get an estimate of the LECs and shape the theoretical results, we employ naive dimensional analysis (NDA)~\cite{Weinberg:1989dx, Manohar:1983md}. This technique does not give an accurate prediction, but it provides order-of-magnitude estimates identifying each source of $P$ and $T$ violation.

NDA estimates are made analogously to~\cite{Unal:2021lhb}. Accordingly, all LECs arising from qEDM and qCEDM are as follows

\begin{equation}
\begin{aligned}
f_{1,2} = & \mathcal O(d_c) =\,\mathcal O\left(\frac{m_c}{\Lambda^2}\right)\,, \label{NDA_qEDM} \\
a_{16\mathrm{-}20} = & \mathcal O\left(e \,\tilde d_c\frac{F_\pi}{ \Lambda_\chi}\right) =\mathcal O\left(e\, \frac{F_\pi m_c}{\Lambda_\chi\Lambda^2}\right)\,,
\end{aligned}
\end{equation}
with $4 \pi F_\pi \sim \Lambda_\chi$. For the 4q and 4qLR operators, LECs emerge from both the tree-level and the $C\!P$-odd interactions at the one-loop level and their NDA representations are
\begin{eqnarray}
\kappa_{6\mathrm{-}10}\, \kappa^{qc}&=& \mathcal O\left(\kappa^{qc} \Lambda_\chi F_\pi^2\right)= \mathcal O \left(\frac{\Lambda_\chi F_\pi^2}{\Lambda^2}\right)\,,\nonumber\\
\kappa_{11\mathrm{-}21}\, \kappa^{qc}&=& \mathcal O\left(e \kappa^{qc} \frac{F_\pi^2}{\Lambda_\chi}\right)= \mathcal O \left(e \frac{F_\pi^2}{\Lambda_\chi \Lambda^2}\right)\,, \nonumber\\
\rho_{1\mathrm{-}5}\, \rho^{qc}&=& \mathcal O\left(\rho^{qc} \Lambda_\chi F_\pi^2\right)= \mathcal O \left(\frac{\Lambda_\chi F_\pi^2}{\Lambda^2}\right)\,,\nonumber\\
\rho_{11\mathrm{-}21}\, \rho^{qc}&=& \mathcal O\left(e \rho^{qc} \frac{F_\pi^2}{\Lambda_\chi}\right)= \mathcal O \left(e \frac{F_\pi^2}{\Lambda_\chi \Lambda^2}\right)\,,
\end{eqnarray}
with $q = \{u,\,d,\,s\}$. In this way, all EDMs are scaled to $\Lambda^{-2}$ so that the sizes of EDMs can be easily obtained for other BSM scales. For a BSM physics scale $\Lambda = 1$ TeV, and considering only the tree-level expressions, the order-of-magnitude estimates of EDMs are
\begin{equation}
d^\gamma_{B_h} = \{10^{-19}\,, 10^{-20}\,,10^{-21}\,,10^{-21}\}\,e\,\mathrm{cm}\,,
\end{equation}
respectively, for the qEDM, qCEDM, 4q, and 4qLR operators. The reasoning behind choosing this scale is explained in~\cite{Unal:2021lhb}. Contrary to the $b$-baryon EDMs calculation, the values of the CKM matrix elements $\text{Re}(V_{dc})$ and $\text{Re}(V_{sc})$ are of order one in the $c$-baryon case.  One can thus immediately notice that the contributions of the 4qLR operator are improved by about three orders of magnitude. These predictions contain a sizeable uncertainty, and thus to get an idea of this sizeable uncertainty determining roughly a range at a given scale $\Lambda$, we employ a Monte Carlo sampling of the LECs. In other words, the LECs for all operators are rescaled and varied the relevant dimensionless constants between $[-3,  +3]$. For the 4q operator, for instance,
	$\kappa_{7} \, (\kappa^{uc} + \kappa^{sc}) \rightarrow \left( \Lambda_{\chi}F_{\pi}^2/\Lambda^2 \right) \tilde{\kappa}_{7} \, ,$
where $ \tilde{\kappa}_{7}$ is the dimensionless constants. This procedure has been done for all LECs appearing in the EDM expression and obtained the ranges for the various EDMs for each $C\!P$-odd source. The results can be read from Tables~\ref{MCEDM-Antitriplet} and~\ref{MCEDM-Sextet}.  While the operator qEDM gives the dominant contribution, 4q and 4qLR terms have the same order of magnitude. As can be seen from the Tables, the EDM predictions for each source are diverse around zero.  On the other hand, the standard deviations are relatively large because of the wide range of the dimensionless constants which was used in the MC sampling.   
One can take a look at the resulting size of the EDM by adding up the single contributions. These values, however, provide an estimate rather than a precise prediction for the range where the EDM of the charmed baryons can be found by future experiments.  For instance, the total EDM of the $\lambda_{c}^+$ from the anti-triplet and $\Omega_{c}^0$ from the sextet baryons would be

\begin{table}[t]
	\centering
	\caption{Contributions to the EDMs of the anti-triplet baryons for $\Lambda = 1\, \text{TeV}$. The results are given in $10^{-19}\,e\,\text{cm}$ and $10^{-20}\,e\,\text{cm}$ for the qEDM and qCEDM operators, respectively, and $10^{-21}\,e\,\text{cm}$ for the 4q and 4qLR operators.}
	\begin{tabular}{lccc}
		\hline
		{Contribution}   &  {$\Lambda_{c}^+$}    & {$\Xi_{c}^0$} &   {$\Xi_{c}^+$}   \\
		\hline  
		{qEDM}  & $-0.14 \pm 1.79$ & $-0.14 \pm 1.79$ & $-0.14 \pm 1.79$ \\
		{qCEDM}  & $-0.10 \pm 3.10 $ & $+0.12 \pm 2.7 $ & $-0.10 \pm 3.10$ \\
		{4q}  & $+0.17 \pm 5.3$ & $+0.17 \pm 4.4$ & $+0.13 \pm 3.9$ \\
		{4qLR}  & $+0.12 \pm 2.6$ & $-0.15 \pm 2.5$ & $-0.11 \pm 2.4$ \\
		\hline
	\end{tabular}
	\label{MCEDM-Antitriplet}
\end{table}
%--

%--
\begin{table}[t]
	\centering
	\caption{Contributions to the EDMs of the sextet baryons for $\Lambda = 1\, \text{TeV}$. The results are given in $10^{-19}\,e\,\text{cm}$ and $10^{-20}\,e\,\text{cm}$ for the qEDM and qCEDM operators, respectively, and $10^{-21}\,e\,\text{cm}$ for the 4q and 4qLR operators.}
	\begin{tabular}{lcccccc}
		\hline
		{Contribution}   &  {$\Sigma_{c}^0$}  & {$\Sigma_{c}^+$}  & {$\Sigma_{c}^{++}$} & {$\Xi_{c}^{'0}$} & {$\Xi_{c}^{'+}$} & {$\Omega_{c}^0$}   \\
		\hline 
		{qEDM}  & $+0.06 \pm 0.86$ & $+0.06 \pm 0.86$ & $+0.06 \pm 0.86$ & $+0.06 \pm 0.86$ & $+0.06 \pm 0.86$ & $+0.06 \pm 0.86$ \\
		{qCEDM}  & $+0.10 \pm 1.37$ & $+0.07 \pm 1.37$ & $-0.13 \pm 1.51$ & $+0.10 \pm 1.37$ & $+0.07 \pm 1.37$ & $+0.10 \pm 1.37$ \\
		{4q}  & $+0.10 \pm 2.90$ & $+0.14 \pm 2.76$ & $+0.28 \pm 4.17$ & $-0.10 \pm 2.32$ & $-0.13 \pm 2.72$ & $+0.19 \pm 2.99$ \\
		{4qLR}  & $+0.01 \pm 0.49$ & $+0.02 \pm 1.37$ & $+0.04 \pm 1.10$ & $+0.03 \pm 1.33$ & $-0.14 \pm 1.58$ & $+0.13 \pm 2.31$ \\
		\hline
	\end{tabular}
	\label{MCEDM-Sextet}
\end{table}
%--

%--
\begin{equation}
\begin{aligned}
d^{\gamma}_{\Lambda_c^+} = & (-0.14 \pm 2.18) \times 10^{-19}\,e\,\text{cm} \, , \\
d^{\gamma}_{\Omega_c^0} = & (0.07 \pm 1.04) \times 10^{-19}\,e\,\text{cm} \, .
\end{aligned}
\end{equation}
The EDM sensitivity of charmed baryons is estimated to be of the order of $10^{-17}\,e\,\text{cm}$  following an experimental scenario considered at the LHC \cite{Bagli:2017foe}. 
Recently, another experiment at the LHC was presented to directly measure the magnetic and electric dipole moments of charmed baryons.
In this work, a new setup was proposed to overcome the difficulties in measuring EDMs with short lifetimes. The expected sensitivity for the EDM of charmed baryons in this ongoing experimental program is of the order of $10^{-16}\,e\,\text{cm}$ \cite{Merli:2022lqf}.
 
\section{Conclusion}
\label{sec:con}

We have analyzed the EDMs of spin-$1/2$ single-charm baryons using the techniques developed in~\cite{Unal:2021lhb} where a similar calculation has been performed for the bottom baryons. $C\!P$-violating effective dimension-6 operators which consist of the so-called quark EDM, the quark-chromo EDM, the 4q operator, and the 4q left-right operator have been compiled in terms of charm and lighter quark bilinears. The resulting $C\!P$-violating hadronic interactions have been constructed between charm baryons, Goldstone bosons, and photons employing the low-energy effective field theory of QCD, heavy baryon chiral perturbation theory (HBChPT ). The EDMs of the charm baryons have been calculated up to the first non-vanishing order for each source of $C\!P$ violation. 

Concerning the qEDM and qCEDM operators, it turns out that patterns of charm baryon EDMs display similar properties with the bottom baryons. However, four-quark operators (4q and 4qLR) display different relations of the EDMs. These patterns can provide the identification of the dominant source of $C\!P$ violation. Moreover, contributions from the 4qLR operator are larger than the $b$-baryon case, as two CKM matrix elements $\text{Re}(V_{dc})$ and $\text{Re}(V_{sc})$ are three orders of magnitude larger than $\text{Re}(V_{ub})$. However, as already pointed out in~\cite{Unal:2021lhb} and refined further above, the absolute sizes of the EDMs are significantly uncertain because of very little information on the LECs in the $C\!P$-odd chiral effective Lagrangian. Therefore, with the contributions for BSM scales of $1$ TeV using NDA assessment, it has been concluded that EDMs are expected in the range of $10^{-19}-10^{-21}$ e cm, which are strongly dependent on chiral- and isospin-symmetry properties of the underlying sources of the dimension-6 operators.  These results and their interpretation call for further studies on both the theoretical and experimental sides. 

\section{Acknowledgements}
The author thanks D. Severt and J. de Vries for helpful discussions and careful reading of the manuscript.
\begin{appendix}
\section{Form Factors}
\label{sec:AppendixA}
Including the tree-level contributions, the full form factor expressions for the $c$-baryons up to the order $\mathcal{O}(\delta^2)$  are 
%--
\begin{equation}
	\begin{aligned} 
	D^{\gamma}_{\Lambda_c^+}(q^2)= ~& 4f_1 + 4e\Big ( a_{16}+ 2 a_{19} + 2 (\kappa_{11}+ 2 \kappa_{20}) \kappa^{uc} + 2 \kappa_{14}(\kappa^{uc}+\kappa^{dc}) \\
	 & + 2  \kappa_{17} (\kappa^{uc}+\kappa^{dc}+\kappa^{sc}) 
	 + 2 \text{Re}(V_{dc})(\rho_{14} +\rho_{17})  \rho^{dc} + 2 \text{Re}(V_{sc}) \rho_{17} \rho^{sc}  \Big)  \\
	      & + \frac{e g_2}{4 F_{\pi}^2}\Big( \text{Re}(V_{sc}) \rho_3 \rho^{sc} +\kappa_8 ( \kappa^{uc}+\kappa^{sc}) \Big) \int_0^1 dx \, \frac{1}{\tilde{M}_K}\frac{\partial}{\partial \tilde{M}_K}J_1( -\Delta, \tilde{M}_K) , 
	      \\
	D^{\gamma}_{\Xi_c^0}(q^2)= ~& 4 f_1 + 8e\Big (a_{19}+\kappa_{14}(\kappa^{dc}+\kappa^{sc}) + 2 \kappa_{20} \kappa^{uc} +( \text{Re}(V_{dc})\rho^{dc} + \text{Re}(V_{sc})\rho^{sc}) \rho_{14} \Big)  \\
		 & - \frac{e g_2}{4 F_{\pi}^2}\Big( \text{Re}(V_{dc}) \rho_3 \rho^{dc} +\kappa_8 ( \kappa^{uc}+\kappa^{dc}) \Big) \int_0^1 dx \, \frac{1}{\tilde{M}_{\pi}}\frac{\partial}{\partial \tilde{M}_{\pi}}J_1(-\Delta, \tilde{M}_{\pi}) 
		 \\
		& - \frac{e g_2}{4 F_{\pi}^2}\Big( \text{Re}(V_{sc}) \rho_3 \rho^{sc} +\kappa_8 ( \kappa^{uc}+\kappa^{sc}) \Big) \int_0^1 dx \, \frac{1}{\tilde{M}_{K}}\frac{\partial}{\partial \tilde{M}_{\pi}}J_1(-\Delta, \tilde{M}_K) ,  
		\\
	D^{\gamma}_{\Xi_c^+}(q^2)=  ~& 4 f_1 + 4e \Big(a_{16}+2a_{19}+2 (\kappa_{11}+ 2 \kappa_{20}) \kappa^{uc} + 2\kappa_{14}(\kappa^{uc}+\kappa^{sc}) \\
	& +2 \kappa_{17} (\kappa^{uc}+\kappa^{dc}+\kappa^{sc}) 
	    + 2 \text{Re}(V_{dc})(\rho_{14} + \rho_{17})\rho^{dc} + 2 \text{Re}(V_{sc}) \rho_{17}\rho^{sc} \Big ) \\
		 & + \frac{e g_2}{4 F_{\pi}^2}\Big( \text{Re}(V_{dc}) \rho_3 \rho^{dc} +\kappa_8 ( \kappa^{uc}+\kappa^{dc}) \Big) \int_0^1 dx \, \frac{1}{\tilde{M}_{\pi}}\frac{\partial}{\partial \tilde{M}_{\pi}}J_1( -\Delta, \tilde{M}_{\pi}) ,   
		 \\
	 D^{\gamma}_{\Sigma_c^0}(q^2)=  ~& 2f_2 + 4e\Big( a_{20} +2 (\kappa_{15} \kappa^{dc}+\kappa_{21} \kappa^{uc}) + 2 \text{Re}(V_{dc})\rho_{15} \rho^{dc} \Big) \\
		 & - \frac{e g_1}{4 F_{\pi}^2}\Big( \text{Re}(V_{dc}) \rho_2 \rho^{dc} +\kappa_7 ( \kappa^{uc}+\kappa^{dc}) \Big) \int_0^1 dx \, \frac{1}{\tilde{M}_{\pi}}\frac{\partial}{\partial \tilde{M}_{\pi}}J_1( -\Delta, \tilde{M}_{\pi}) 
	    \\
		& - \frac{e g_2}{2 F_{\pi}^2}\Big( \text{Re}(V_{dc}) \rho_3 \rho^{dc} +\kappa_8 ( \kappa^{uc}+\kappa^{dc})  \Big) \int_0^1 dx \, \frac{1}{\tilde{M}_{\pi}}\frac{\partial}{\partial \tilde{M}_{\pi}}J_1(0, \tilde{M}_{\pi}) ,  \\
	D^{\gamma}_{\Sigma_c^+}(q^2)=  ~&  2 f_2 + e \Big( a_{17}+4a_{20}+ 4(\kappa_{12}+ 2 \kappa_{21}) \kappa^{uc} +4 \kappa_{15}(\kappa^{uc}+\kappa^{dc}) \\
	 &+ 4\kappa_{18} (\kappa^{uc}+\kappa^{dc}+\kappa^{sc})
        + 4 \text{Re}(V_{dc})(\rho_{15} +\rho_{18})  \rho^{dc} + 4\text{Re}(V_{sc}) \rho_{18} \rho^{sc} \Big ) \\
	    & + \frac{e g_1}{8 F_{\pi}^2}\Big( \text{Re}(V_{sc}) \rho_2 \rho^{sc} +\kappa_7 ( \kappa^{uc}+\kappa^{sc}) \Big) \int_0^1 dx \, \frac{1}{\tilde{M}_{K}}\frac{\partial}{\partial \tilde{M}_{K}}J_1( -\Delta, \tilde{M}_{K}) 
	    \\
		& + \frac{e g_2}{4 F_{\pi}^2}\Big( \text{Re}(V_{sc}) \rho_3 \rho^{sc} +\kappa_8 ( \kappa^{uc}+\kappa^{sc})  \Big) \int_0^1 dx \, \frac{1}{\tilde{M}_{K}}\frac{\partial}{\partial \tilde{M}_{K}}J_1(0, \tilde{M}_{K}) , 
	\end{aligned}
\end{equation} 
%--

%--
\begin{equation}
	\vspace{85pt}
	\begin{aligned} 
	D^{\gamma}_{\Sigma_c^{++}}(q^2)=  ~& 2 f_2 + 2e \Big(a_{17}+2a_{20}+4 (\kappa_{12}+ \kappa_{15}+ \kappa_{21}) \kappa^{uc} + 4\kappa_{18}(\kappa^{uc}+\kappa^{dc}+\kappa^{sc}) 
	  \\
	     &+ 4 \big(\text{Re}(V_{dc})\rho^{dc}+\text{Re}(V_{sc})\rho^{sc}\big) \rho_{18} \Big) \\ 
		& + \frac{e g_1}{4 F_{\pi}^2}\Big( \text{Re}(V_{dc}) \rho_2 \rho^{dc} +\kappa_7 ( \kappa^{uc}+\kappa^{dc}) \Big)  \int_0^1 dx \, \frac{1}{\tilde{M}_{\pi}}\frac{\partial}{\partial \tilde{M}_{\pi}}J_1( -\Delta, \tilde{M}_{\pi})
		 \\
		& + \frac{e g_1}{4 F_{\pi}^2}\Big( \text{Re}(V_{sc}) \rho_2 \rho^{sc} +\kappa_7 ( \kappa^{uc}+\kappa^{sc}) \Big)  \int_0^1 dx \, \frac{1}{\tilde{M}_{K}}\frac{\partial}{\partial \tilde{M}_{K}}J_1( -\Delta, \tilde{M}_{K}) 
		\\
		& + \frac{e g_2}{2 F_{\pi}^2} \Big( \text{Re}(V_{dc}) \rho_3 \rho^{dc} +\kappa_8 ( \kappa^{uc}+\kappa^{dc}) \Big)  \int_0^1 dx \, \frac{1}{\tilde{M}_{\pi}}\frac{\partial}{\partial \tilde{M}_{\pi}}J_1( 0, \tilde{M}_{\pi})
		\\
		& + \frac{e g_2}{2 F_{\pi}^2} \Big( \text{Re}(V_{sc}) \rho_3 \rho^{sc} +\kappa_8 ( \kappa^{uc}+\kappa^{sc}) \Big)  \int_0^1 dx \, \frac{1}{\tilde{M}_{K}}\frac{\partial}{\partial \tilde{M}_{K}}J_1( 0, \tilde{M}_{K}) ,  
		\\
		D^{\gamma}_{\Xi_c^{'0}}(q^2)=  ~&  2 f_2 + 2e \Big( 2 a_{20} + 2 \kappa_{15}(\kappa^{dc}+\kappa^{sc}) + 4\kappa_{21} \kappa^{uc} 
		 + 2 \big(\text{Re}(V_{dc}) \rho^{dc}+ \text{Re}(V_{sc}) \rho^{sc}\big) \rho_{15}  \Big) \\
		& - \frac{e g_1}{8 F_{\pi}^2}\Big( \text{Re}(V_{dc}) \rho_2 \rho^{dc} +\kappa_7 ( \kappa^{uc}+\kappa^{dc}) \Big)  \int_0^1 dx \, \frac{1}{\tilde{M}_{\pi}}\frac{\partial}{\partial \tilde{M}_{\pi}}J_1( -\Delta, \tilde{M}_{\pi})
		 \\
		& - \frac{e g_1}{8 F_{\pi}^2}\Big( \text{Re}(V_{sc}) \rho_2 \rho^{sc} +\kappa_7 ( \kappa^{uc}+\kappa^{sc}) \Big)  \int_0^1 dx \, \frac{1}{\tilde{M}_{K}}\frac{\partial}{\partial \tilde{M}_{K}}J_1( -\Delta, \tilde{M}_{K}) 
		\\
		& -  \frac{e g_2}{4 F_{\pi}^2} \Big( \text{Re}(V_{dc}) \rho_3 \rho^{dc} +\kappa_8 ( \kappa^{uc}+\kappa^{dc}) \Big)  \int_0^1 dx \, \frac{1}{\tilde{M}_{\pi}}\frac{\partial}{\partial \tilde{M}_{\pi}}J_1( 0, \tilde{M}_{\pi})
		\\
		& - \frac{e g_2}{4 F_{\pi}^2} \Big( \text{Re}(V_{sc}) \rho_3 \rho^{sc} +\kappa_8 ( \kappa^{uc}+\kappa^{sc}) \Big)  \int_0^1 dx \, \frac{1}{\tilde{M}_{K}}\frac{\partial}{\partial \tilde{M}_{K}}J_1( 0, \tilde{M}_{K}) ,  
		\\
	  D^{\gamma}_{\Xi_c^{'+}}(q^2)=  ~& 2 f_2 + e \Big( a_{17} + 4a_{20} + 4 (\kappa_{12}+ 2 \kappa_{21}) \kappa^{uc} + 4 \kappa_{15}(\kappa^{uc}+\kappa^{sc}) \\
	   &+ 4\kappa_{18} (\kappa^{uc}+\kappa^{dc}+\kappa^{sc})
	   +4 \text{Re}(V_{sc})(\rho_{15} +\rho_{18})  \rho^{sc} + 4 \text{Re}(V_{dc}) \rho_{18} \rho^{dc} \Big) \\
		& + \frac{e g_1}{8 F_{\pi}^2}\Big( \text{Re}(V_{dc}) \rho_2 \rho^{dc} +\kappa_7 ( \kappa^{uc}+\kappa^{dc}) \Big) \int_0^1 dx \, \frac{1}{\tilde{M}_{\pi}}\frac{\partial}{\partial \tilde{M}_{\pi}}J_1( -\Delta, \tilde{M}_{\pi}) \\
		& + \frac{e g_2}{4 F_{\pi}^2} \Big( \text{Re}(V_{dc}) \rho_3 \rho^{dc}+\kappa_8 ( \kappa^{uc}+\kappa^{dc}) \Big) \int_0^1 dx \, \frac{1}{\tilde{M}_{\pi}}\frac{\partial}{\partial \tilde{M}_{\pi}}J_1( 0, \tilde{M}_{\pi}) , \\
      D^{\gamma}_{\Omega_c^0}(q^2)=  ~& 2 f_2 +4 e\Big( a_{20} +  2\kappa_{21}\kappa^{uc} + 2 \kappa_{15}\kappa^{sc}+2 \text{Re}(V_{sc}) \rho_{15} \rho^{sc}\Big) \\
		& - \frac{e g_1}{4 F_{\pi}^2}\Big( \text{Re}(V_{sc}) \rho_2 \rho^{sc} +\kappa_7 ( \kappa^{uc}+\kappa^{sc}) \Big)  \int_0^1 dx \, \frac{1}{\tilde{M}_K}\frac{\partial}{\partial \tilde{M}_K}J_1( -\Delta, \tilde{M}_K) \\
		& - \frac{e g_2}{2 F_{\pi}^2} \Big( \text{Re}(V_{sc}) \rho_3 \rho^{sc} +\kappa_8 ( \kappa^{uc}+\kappa^{sc}) \Big) \int_0^1 dx \, \frac{1}{\tilde{M}_K}\frac{\partial}{\partial \tilde{M}_K}J_1( 0, \tilde{M}_K) . \nonumber
			\end{aligned}
	\label{all-edm-form}
\end{equation}
%--

\section{EDM Estimates with Naive Dimensional Analysis}
\label{sec:AppendixB}
In this appendix, we give $c$ baryon EDMs with the NDA estimate. Replacement of the unknown low energy constants with their NDA correspondence yield 
%--
\begin{equation}
\begin{aligned} 
d^{\gamma}_{\Lambda_c^+}= ~& \frac{4 m_c \tilde{f}_1  }{\Lambda^2} +  \frac{e m_c}{ \pi \Lambda^2} (\tilde{a}_{16}+2\tilde{a}_{19}) \\
     &+  \frac{2 e F_{\pi}}{ \pi \Lambda^2} \Big( \tilde{\kappa}_{11} +  \tilde{\kappa}_{14} + \tilde{\kappa}_{17} + 2 \tilde{\kappa}_{20} 
     +  \text{Re}(V_{dc}) ( \tilde{\rho}_{14} + \tilde{\rho}_{17}) + \text{Re}(V_{sc}) \tilde{\rho}_{17}  \Big) \\
     & + \frac{e g_2 \Lambda_\chi}{32 \pi^2 \Lambda^2} \Big(  \text{Re}(V_{sc}) \tilde{\rho}_{3} + \tilde{\kappa}_{8} \Big) 
		\Bigg( 1 + 2 \text{ln}\Bigg[\frac{M_K}{\lambda}\Bigg] + \frac{2 \Delta \text{Arccos}\big[\frac{\Delta}{M_K}\big]}{\sqrt{M_K^2-\Delta^2}} \Bigg), 
		\\                                  
d^{\gamma}_{\Xi_c^0}= ~& \frac{4 m_c  \tilde{f}_1  }{\Lambda^2} +  \frac{2 e m_c }{ \pi \Lambda^2} \tilde{a}_{19} +  \frac{2 e F_{\pi}}{ \pi \Lambda^2} \Bigg( \tilde{\kappa}_{14} +  2 \tilde{\kappa}_{20} 
     + \Big( \text{Re}(V_{dc}) + \text{Re}(V_{sc})\Big)  \tilde{\rho}_{14} \Bigg)
     \\
		& - \frac{e g_2 \Lambda_\chi}{32 \pi^2 \Lambda^2} \Big(  \text{Re}(V_{dc}) \tilde{\rho}_{3} + \tilde{\kappa}_{8} \Big) \Bigg( 1 + 2 \text{ln}\Bigg[\frac{M_{\pi}}{\lambda}\Bigg] + \frac{2 \Delta}{\sqrt{\Delta^2-M_{\pi}^2}} \text{ln}\Bigg[ \frac{\Delta}{M_{\pi}}  + \sqrt{\frac{\Delta^2}{M_{\pi}^2} -1 } \Bigg] \Bigg)
		\\     
		& - \frac{e g_2 \Lambda_\chi}{32 \pi^2 \Lambda^2} \Big(  \text{Re}(V_{sc}) \tilde{\rho}_{3} + \tilde{\kappa}_{8} \Big) 
		\Bigg( 1 + 2 \text{ln}\Bigg[\frac{M_K}{\lambda}\Bigg] + \frac{2 \Delta \text{Arccos}\big[\frac{\Delta}{M_K}\big]}{\sqrt{M_K^2-\Delta^2}} \Bigg),      
		\\        
d^{\gamma}_{\Xi_c^+} = ~& \frac{4 m_c \tilde{f}_1 }{\Lambda^2} +  \frac{e m_c}{ \pi \Lambda^2} (\tilde{a}_{16}+2\tilde{a}_{19}) \\
   & +  \frac{2 e F_{\pi}}{ \pi \Lambda^2} \Big( \tilde{\kappa}_{11} +   \tilde{\kappa}_{14} + \tilde{\kappa}_{17} +  \tilde{\kappa}_{20} 
     +  \text{Re}(V_{dc}) ( \tilde{\rho}_{14} + \tilde{\rho}_{17}) +  \text{Re}(V_{sc}) \tilde{\rho}_{17}  \Big)  \\
     & + \frac{e g_2 \Lambda_\chi }{32 \pi^2 \Lambda^2} \Big(  \text{Re}(V_{dc}) \tilde{\rho}_{3} + \tilde{\kappa}_{8} \Big)
 \Bigg( 1+ 2 \text{ln}\Bigg[\frac{M_{\pi}}{\lambda}\Bigg] + \frac{2 \Delta}{\sqrt{\Delta^2-M_{\pi}^2}} \text{ln}\Bigg[ \frac{\Delta}{M_{\pi}}  + \sqrt{\frac{\Delta^2}{M_{\pi}^2} -1 } \Bigg]  \Bigg), 
		\\ 
d^{\gamma}_{\Sigma_c^0}=  ~&  \frac{2 m_c \tilde{f}_2 }{\Lambda^2}  + \frac{e m_c }{ \pi \Lambda^2} \tilde{a}_{20}  +  \frac{2 e F_{\pi}}{ \pi \Lambda^2} \Big( \tilde{\kappa}_{15} + \tilde{\kappa}_{21} + \text{Re}(V_{dc})  \tilde{\rho}_{15} \Big)
\\ 
		& -\frac{e g_1}{32 \pi^2} \frac{\Lambda_\chi}{\Lambda^2} \Big(  \text{Re}(V_{dc}) \tilde{\rho}_{2} + \tilde{\kappa}_{7} \Big) 
		\Bigg( 1 + 2 \text{ln}\Bigg[\frac{M_{\pi}}{\lambda}\Bigg] + \frac{2 \Delta}{\sqrt{\Delta^2-M_{\pi}^2}} \text{ln}\Bigg[ \frac{\Delta}{M_{\pi}}  + \sqrt{\frac{\Delta^2}{M_{\pi}^2} -1 } \Bigg] \Bigg)
		\\
		& - \frac{e g_2}{16 \pi^2}\frac{\Lambda_\chi}{\Lambda^2} \Big( \text{Re}(V_{dc}) \tilde{\rho}_{3} + \tilde{\kappa}_{8} \Big) \Bigg(1 + 2 \text{ln} \Bigg[\frac{M_\pi}{\lambda}\Bigg]\Bigg),  \\
		d^{\gamma}_{\Sigma_c^+} = ~& \frac{2 m_c \tilde{f}_2 }{\Lambda^2} +  \frac{e m_c}{ 4 \pi \Lambda^2} (\tilde{a}_{17}+4\tilde{a}_{20}) \\
		& +  \frac{e F_{\pi}}{ \pi \Lambda^2} \Big( \tilde{\kappa}_{12} +   \tilde{\kappa}_{15} + \tilde{\kappa}_{18} +  2\tilde{\kappa}_{21} 
     +  \text{Re}(V_{dc}) ( \tilde{\rho}_{15} + \tilde{\rho}_{18}) +  \text{Re}(V_{sc}) \tilde{\rho}_{18}  \Big) \\
        & + \frac{e g_1}{64 \pi^2} \frac{\Lambda_\chi}{\Lambda^2} \Big(  \text{Re}(V_{sc}) \tilde{\rho}_{2} + \tilde{\kappa}_{7} \Big)
		 \Bigg( 1 + 2 \text{ln}\Bigg[\frac{M_K}{\lambda}\Bigg] + \frac{2 \Delta \text{Arccos}\big[\frac{\Delta}{M_K}\big]}{\sqrt{M_K^2-\Delta^2}}  \Bigg) 
		 \\
		 &- \frac{e g_2}{32 \pi^2}\frac{\Lambda_\chi}{\Lambda^2} \Big( \text{Re}(V_{sc}) \tilde{\rho}_{3} + \tilde{\kappa}_{8} \Big) \Bigg(1 + 2 \text{ln} \Bigg[\frac{M_K}{\lambda}\Bigg]\Bigg), 
\label{all-edm-nda}
\end{aligned}
\end{equation} 

\begin{equation}
\vspace{10pt}
\begin{aligned} 
d^{\gamma}_{\Sigma_c^{++}} = ~&  \frac{2 m_c \tilde{f}_2 }{\Lambda^2} +  \frac{e m_c}{ 2 \pi \Lambda^2} (\tilde{a}_{17}+2\tilde{a}_{20}) +  \frac{2e F_{\pi}}{ \pi \Lambda^2} \Big( \tilde{\kappa}_{12} +   \tilde{\kappa}_{15} + \tilde{\kappa}_{18} +  \tilde{\kappa}_{21} 
     + \big( \text{Re}(V_{dc}) + \text{Re}(V_{sc})\big) \tilde{\rho}_{18}  \Big)
 \\ 
		 &+ \frac{e g_1}{32 \pi^2} \frac{\Lambda_\chi}{\Lambda^2} \Big(  \text{Re}(V_{dc}) \tilde{\rho}_{2} + \tilde{\kappa}_{7} \Big) \Bigg( 1  + 2 \text{ln}\Bigg[\frac{M_{\pi}}{\lambda}\Bigg] + \frac{2 \Delta}{\sqrt{\Delta^2-M_{\pi}^2}} \text{ln}\Bigg[ \frac{\Delta}{M_{\pi}}  + \sqrt{\frac{\Delta^2}{M_{\pi}^2} -1 } \Bigg] \Bigg) 
				\\
			 &+ \frac{e g_1}{32 \pi^2} \frac{\Lambda_\chi}{\Lambda^2} \Big(  \text{Re}(V_{sc}) \tilde{\rho}_{2} + \tilde{\kappa}_{7} \Big)\Bigg( 1 + 2 \text{ln}\Bigg[\frac{M_K}{\lambda}\Bigg] + \frac{2 \Delta \text{Arccos}\big[\frac{\Delta}{M_K}\big]}{\sqrt{M_K^2-\Delta^2}}  \Bigg)  	\\	
		& + \frac{e g_2}{16 \pi^2} \frac{\Lambda_\chi}{\Lambda^2} \Big(  \text{Re}(V_{dc}) \tilde{\rho}_{3} + \tilde{\kappa}_{8} \Big) \Bigg( 1 + 2 \text{ln}\Bigg[\frac{ M_{\pi} }{\lambda}\Bigg] \Bigg) 
		\\
		& + \frac{e g_2}{16 \pi^2} \frac{\Lambda_\chi}{\Lambda^2} \Big(  \text{Re}(V_{sc}) \tilde{\rho}_{3} + \tilde{\kappa}_{8} \Big) \Bigg( 1 + 2 \text{ln}\Bigg[\frac{ M_{K} }{\lambda}\Bigg] \Bigg) ,
		\\
		d^{\gamma}_{\Xi_c^{'0}}= ~&  \frac{2 m_c \tilde{f}_2 }{\Lambda^2} +  \frac{e m_c}{ \pi \Lambda^2} \tilde{a}_{20} +  \frac{e F_{\pi}}{ \pi \Lambda^2} \Big( \tilde{\kappa}_{15} + 2 \tilde{\kappa}_{21} 
     + \big( \text{Re}(V_{dc}) + \text{Re}(V_{sc})\big) \tilde{\rho}_{15}  \Big)
 \\ 
		 &- \frac{e g_1}{64 \pi^2} \frac{\Lambda_\chi}{\Lambda^2} \Big(  \text{Re}(V_{dc}) \tilde{\rho}_{2} + \tilde{\kappa}_{7} \Big) \Bigg( 1  + 2 \text{ln}\Bigg[\frac{M_{\pi}}{\lambda}\Bigg] + \frac{2 \Delta}{\sqrt{\Delta^2-M_{\pi}^2}} \text{ln}\Bigg[ \frac{\Delta}{M_{\pi}}  + \sqrt{\frac{\Delta^2}{M_{\pi}^2} -1 } \Bigg] \Bigg) 	\\
			 &- \frac{e g_1}{64 \pi^2} \frac{\Lambda_\chi}{\Lambda^2} \Big(  \text{Re}(V_{sc}) \tilde{\rho}_{2} + \tilde{\kappa}_{7} \Big)\Bigg( 1 + 2 \text{ln}\Bigg[\frac{M_K}{\lambda}\Bigg] + \frac{2 \Delta \text{Arccos}\big[\frac{\Delta}{M_K}\big]}{\sqrt{M_K^2-\Delta^2}}  \Bigg) 
				\\
		& - \frac{e g_2}{32 \pi^2} \frac{\Lambda_\chi}{\Lambda^2} \Big(  \text{Re}(V_{dc}) \tilde{\rho}_{3} + \tilde{\kappa}_{8} \Big) \Bigg( 1 + 2 \text{ln}\Bigg[\frac{ M_{\pi} }{\lambda}\Bigg] \Bigg) 
		\\
		& - \frac{e g_2}{32 \pi^2} \frac{\Lambda_\chi}{\Lambda^2} \Big(  \text{Re}(V_{sc}) \tilde{\rho}_{3} + \tilde{\kappa}_{8} \Big) \Bigg( 1 + 2 \text{ln}\Bigg[\frac{ M_{K} }{\lambda}\Bigg] \Bigg) ,
		\\
		d^{\gamma}_{\Xi_c^{'+}} = ~& \frac{2 m_c \tilde{f}_2 }{\Lambda^2} +  \frac{e m_c}{ 4 \pi \Lambda^2} (\tilde{a}_{17}+4\tilde{a}_{20}) \\
		 &+  \frac{e F_{\pi}}{ \pi \Lambda^2} \Big( \tilde{\kappa}_{12} +   \tilde{\kappa}_{15} + \tilde{\kappa}_{18} +  2\tilde{\kappa}_{21} 
     + \big( \text{Re}(V_{sc}) (\tilde{\rho}_{15}+\tilde{\rho}_{18}) + \text{Re}(V_{dc})\tilde{\rho}_{18}\big)  \Big)
 \\ 
				& + \frac{e g_1}{64 \pi^2} \frac{\Lambda_\chi}{\Lambda^2} \Big(  \text{Re}(V_{dc}) \tilde{\rho}_{2} + \tilde{\kappa}_{7} \Big)  \Big) \Bigg( 1  + 2 \text{ln}\Bigg[\frac{M_{\pi}}{\lambda}\Bigg] + \frac{2 \Delta}{\sqrt{\Delta^2-M_{\pi}^2}} \text{ln}\Bigg[ \frac{\Delta}{M_{\pi}}  + \sqrt{\frac{\Delta^2}{M_{\pi}^2} -1 } \Bigg] \Bigg) 
				\\
				& + \frac{e g_2}{64 \pi^2} \frac{\Lambda_\chi}{\Lambda^2} \Big(  \text{Re}(V_{sc}) \tilde{\rho}_{3} + \tilde{\kappa}_{8} \Big) \Bigg( 1 + 2 \text{ln}\Bigg[\frac{ M_{\pi} }{\lambda}\Bigg] \Bigg) , \\
		d^{\gamma}_{\Omega_c^0} = ~& \frac{2 m_c \tilde{f}_2 }{\Lambda^2} +  \frac{e m_c}{ \pi \Lambda^2} \tilde{a}_{20} +  \frac{2e F_{\pi}}{ \pi \Lambda^2} \Big( \tilde{\kappa}_{15} +  \tilde{\kappa}_{21} 
     + \text{Re}(V_{sc}) \tilde{\rho}_{15}  \Big)
 \\ 
				& - \frac{e g_1}{32 \pi^2} \frac{\Lambda_\chi}{\Lambda^2} \Big( \text{Re}(V_{sc}) \tilde{\rho}_{2} + \tilde{\kappa}_{7} \Big) \Bigg( 1 + 2 \text{ln}\Bigg[\frac{M_K}{\lambda}\Bigg] + \frac{2 \Delta \text{Arccos}\big[\frac{\Delta}{M_K}\big]}{\sqrt{M_K^2-\Delta^2}} \Bigg) \\
				& - \frac{e g_2}{16 \pi^2} \frac{\Lambda_\chi}{\Lambda^2} \Big(  \text{Re}(V_{sc}) \tilde{\rho}_{3} + \tilde{\kappa}_{8} \Big) \Bigg( 1 + 2 \text{ln}\Bigg[\frac{ M_K }{\lambda}\Bigg] \Bigg) ,\nonumber
	\end{aligned}
	\label{all-edm-nda}
\end{equation}
%--
where the estimation $4 \pi F_\pi \sim \Lambda_\chi$ has been utilized, and the dimensionless constants $\tilde{f}_i$, $\tilde{a}_i$, $\tilde{\kappa}_i$, and $\tilde{\rho}_i$ are varied from $-3$ to $+3$.

\section{Loop Functions}
\label{sec:AppendixC}
In the heavy-baryon formulation \cite{Bernard:1995dp}, the loop functions 
which was used in the calculation of the diagrams in Figure~\ref{fig:diag} are given as
%--
\begin{equation}
\begin{aligned}
    \Delta_{M}=~& 2 M^2 \, \Bigg[ L + \frac{1}{16 \pi^2} \text{ln}\Bigg(\frac{M}{\lambda}\Bigg)\Bigg]+\mathcal{O}(n-4), \\
   \frac{1}{i} \bigintssss \frac{d^n k}{(2 \pi)^n} & \frac{1}{M^2-k^2} = ~\Delta_{M}=M^{n-2}\,(4 \pi)^{-n/2}\, \Gamma \Big(1-\frac{n}{2}\Big),
\end{aligned}
\end{equation}	
%--
\begin{equation}
	 \frac{1}{i} \bigintssss \frac{d^n k}{(2 \pi)^n} \frac{\{1,k_{\mu}, k_{\mu} k_{ \nu}\}}{[v \cdot k-w][M^2-k^2]}=\Big\{J_0(w), v_{\mu} J_1(w), g_{\mu \nu} J_2(w)+v_{\mu} v_{\nu} J_3(w)\Big\} ,
\end{equation}
%--
\begin{equation}
	\frac{1}{i} \bigintssss \frac{d^n k}{(2 \pi)^n} \frac{\{1,k_{\mu}, k_{\mu} k_{ \nu}\}}{[v \cdot k-w]^2 [M^2-k^2]} =  \Big\{G_0(w), v_{\mu} G_1(w), g_{\mu \nu} G_2(w)+v_{\mu} v_{\nu} G_3(w)\Big\} ,
\end{equation}
%--
\begin{equation}
	\frac{1}{i} \bigintssss \frac{d^n k}{(2 \pi)^n}  \frac{1}{[v \cdot k-w)[M^2-k^2][(k+q)^2-M^2]} =
	\bigintssss_0^1 dx \, \frac{1}{2 \widetilde{M}} \, \frac{\partial}{\partial \widetilde{M}} \, J_0 \Big(\widetilde{w}, \widetilde{M}\Big) ,
\end{equation}
%--
\begin{equation}
\begin{aligned}
	   \frac{1}{i} \bigintssss \frac{d^n k}{(2 \pi)^n}  \frac{k_{\mu}}{[v \cdot k-w][M^2-k^2][(k+q)^2-M^2]}  = \\
	   \bigintssss_0^1 dx \, \Bigg(\frac{ v_{\mu}}{2 \widetilde{M}} \, \frac{\partial}{\partial \widetilde{M}} \, J_1 \Big(\widetilde{w}, \widetilde{M}\Big) 
	                                   - \frac{x \, q_{\mu}}{2 \widetilde{M}} \, \frac{\partial}{\partial \widetilde{M}} \, J_0 \Big(\widetilde{w}, \widetilde{M}\Big)\Bigg) ,
\end{aligned}
\end{equation}
%--
\begin{equation}
	\begin{aligned}
	 & \frac{1}{i} \bigintssss \frac{d^n k}{(2 \pi)^n} \frac{k_{\mu} k_{ \nu}}{[v \cdot k-w][M^2-k^2][(k+q)^2-M^2]} = \\
						  & \bigintssss_0^1 dx \Bigg(\frac{ g_{\mu \nu}}{2 \widetilde{M}} \, \frac{\partial}{\partial \widetilde{M}} \, J_2 \Big(\widetilde{w}, \widetilde{M}\Big) + \frac{v_{\mu} v_{\nu}}{2 \widetilde{M}} \, \frac{\partial}{\partial \widetilde{M}}\, J_3 \Big(\widetilde{w}, \widetilde{M}\Big) \\
						& - \left(  q_{\mu} v_{\nu} + q_{\nu} v_{\mu} \right)  \frac{x}{2 \widetilde{M}} \, \frac{\partial}{\partial \widetilde{M}} \, J_1 \Big(\widetilde{w}, \widetilde{M}\Big) 
						   + \frac{x^2 \, q_{\mu} q_{\nu}}{2 \widetilde{M}} \frac{\partial}{\partial \widetilde{M}} \,J_0 \Big(\widetilde{w}, \widetilde{M}\Big) \Bigg) ,  
   \end{aligned}
\end{equation}
%--
where $\widetilde{w}(x)= w + x v \cdot q$, and $\widetilde{M}^2(x)=x(x-1)q^2+M^2$.  In dimensional regularization, the analytical expressions for the loop functions are 
%--
\begin{equation}
	J_0(w)=~-4Lw+\frac{w}{8 \pi^2}\Bigg[1-2 \, \text{ln}\Bigg(\frac{M}{\lambda}\Bigg)\Bigg]-\frac{1}{4 \pi^2} \sqrt{M^2-w^2} \, \text{ArcCos}\Big(\frac{-w}{M}\Big)+\mathcal{O}(n-4) ,
\end{equation}
%--
for $M^2>w^2$, 
%--
\begin{equation}
J_0(w)=~-4Lw+\frac{w}{8 \pi^2}\Bigg[1-2 \, \text{ln}\Bigg(\frac{M}{\lambda}\Bigg)\Bigg]+\frac{1}{4 \pi^2} \sqrt{w^2-M^2} \, \text{ln}\Bigg(\frac{-w}{M} + \sqrt{ \frac{w^2}{M^2} -1} \Bigg)+\mathcal{O}(n-4) ,
\end{equation}
%--
for $w^2>M^2$, and 
%--
\begin{equation}
J_1(w)=~ wJ_0(w)+\Delta_M, \quad J_2(w)=\frac{1}{n-1} \Bigg[(M^2-w^2) J_0(w)-w\Delta_M \Bigg] ,
\end{equation}
%--	
\begin{equation}
J_3(w)=~ wJ_1(w)-J_2(w) ,
\end{equation}
%--
\begin{equation}
G_i(w)=~ \frac{\partial}{\partial w} J_i(w), \quad i=0,1,2,3.
\end{equation}
%--
\end{appendix}

\end{document}